\documentclass[aps,pra,showpacs]{revtex4}
\usepackage{graphicx}
\usepackage{subfigure}
\usepackage{caption2}
\usepackage{float}
\usepackage{amsmath}
\usepackage{amssymb}

\begin{document}

\title{Two-component gap solitons in two- and one-dimensional Bose-Einstein
condensates}
\author{A. Gubeskys, B. A. Malomed, and I. M. Merhasin \\
Department of Interdisciplinary Studies, School of Electrical Engineering\\
Faculty of Engineering, Tel Aviv University\\
Tel Aviv 69978, Israel}

\begin{abstract}
We introduce two- and one-dimensional (2D and 1D) models of a
binary BEC (Bose-Einstein condensate) in a periodic potential,
with repulsive interactions. We chiefly consider the most
fundamental case of the inter-species repulsion with zero
intra-species interactions. The same system may also model a
mixture of two mutually repulsive fermionic species. Existence and
stability regions for gap solitons (GSs) supported by the
interplay of the inter-species repulsion and periodic potential
are identified. Two-component GSs are constructed by means of the
variational approximation (VA) and in a numerical form. The VA
provides accurate description for the GS which is a bound state of
two tightly-bound components, each essentially trapped in one cell
of the periodic potential. GSs of this type dominate in the case
of \textit{intra-gap} solitons, with both components belonging to
the first finite bandgap of the linear spectrum (only this type of
solitons is possible in a weak lattice). \textit{Inter-gap}
solitons, with one component residing in the second bandgap, and
intra-gap solitons which have both components in the second gap,
are possible in a deeper periodic potential, with the strength
essentially exceeding the recoil energy of the atoms. Inter-gap
solitons are, typically, bound states of one tightly- and one
loosely-bound components. In this case, results are obtained in a
numerical form. The number of atoms in experimentally relevant
situations is estimated to be $\sim 5,000$ in 2D intra-gap
soliton, and $\sim 25,000$ in its inter-gap counterpart; in 1D
solitons, it may be up to $10^{5}$. For 2D solitons, the stability
is identified in direct simulations, while in the 1D case it is
done via eigenfrequencies of small perturbations, and then
verified by simulations. In the latter case, if the intra-gap
soliton in the first bandgap is weakly unstable, it evolves into a
stable breather, while unstable solitons of other types (in
particular, inter-gap solitons) get completely destroyed. The
intra-gap 2D solitons in the first bandgap are less robust, and in
some cases they are completely destroyed by the instability.
Addition of intra-species repulsion to the repulsion between the
components leads to further stabilization of the GSs.
\end{abstract}

\maketitle

\section{Introduction}

Solitons in Bose-Einstein condensates (BECs) have drawn a great
deal of attention as robust nonlinear matter-wave pulses. Bright
(localized) solitons were first experimentally created in an
effectively one-dimensional (1D) condensate of $^{7}$Li, loaded in
a strongly elongated (nearly one-dimensional) ``cigar-shaped" trap
\cite{Li}. The use of the Feshbach resonance (FR) made it possible
to keep the scattering length in the condensate negative, with a
very small absolute value, $\sim 0.1$ nm. The weak self-attractive
nonlinearity controlled by means of this technique was sufficient
to create 1D solitons, being a way below the collapse threshold in
the cigar-shape configuration, thus securing the stability of the
solitons.

More generic for BEC is a positive scattering length,
corresponding to repulsive interactions between atoms. In this
case, bright solitons may be created as a result of the interplay
of the intrinsic repulsion and periodic potential induced by an
optical lattice (OL, i.e., the interference pattern created by
counterpropagating beams illuminating the condensate). It was
predicted \cite{GS,Ostrovskaya} that \textit{gap solitons} (GSs)
may emerge in bandgaps of the system's spectrum, since the
interplay of a \emph{negative} effective mass, appearing in a part
of the gap, with the repulsive interaction is exactly what is
needed to create a bright soliton. Theoretical models for GSs in
BEC were reviewed in Ref. \cite{Konotop}, and a rigorous stability
analysis for them was developed in Ref. \cite{Pelinovsky}.
Experimentally, a GS in the $^{87}$Rb condensate with a positive
scattering length, loaded into a cigar-shaped trap equipped with a
longitudinal OL, was for the first time created in Ref.
\cite{Oberthaler} (the soliton was composed of $\sim 1000$ atoms).

Binary mixtures of BECs are also available to experimental
studies. Most typically, they contain two different hyperfine
states of the same atomic species, such as $^{87}$Rb \cite{Rb} and
$^{23}$Na \cite{Na}. BEC was also created in a
\textit{heteronuclear} mixture of $^{41}$K and $^{87}$Rb
\cite{hetero}. As mentioned above, the magnitude and sign of the
scattering lengths of collisions between atoms in the same species
may be altered, via the FR technique, by an external spatially
uniform magnetic \cite{Feshbach} or optical
\cite{optical-Feshbach} field. The scattering length of collisions
between atoms belonging to different species may also be
controlled by the magnetic field \cite{inter-Feshbach}.

The latter possibility opens a way to create a binary mixture with
the (natural) intra-species repulsion, while the sign of the
inter-species interaction is switched to attraction. In recent
preprints \cite{symbiotic}, it was proposed to use this
possibility to create \textit{symbiotic} bright solitons: while
each self-repulsive species cannot support a soliton by itself,
the inter-species attraction opens a way to make two-component
solitons. A somewhat similar possibility was earlier proposed in
terms of a Bose-Fermi mixture, where the interaction between
bosons is repulsive, but the bosons and fermions attract each
other \cite{Poland}. Related to the latter setting, is a proposal
to use attraction between fermions and bosons to built bosonic
quantum dots for fermions (in particular, gap solitons in the BEC
trapped in an OL may play the role of such quantum dots)
\cite{Salerno}.

In the present work, the aim to construct 2D and 1D solutions for
two-component GSs in the most natural setting, when the inter-species
interaction is repulsive. We will chiefly focus on the basic case, when
intra-species interactions may be completely neglected, while the interplay
of the OL potential and repulsion between the two species help to build GSs.
The actual possibility to nullify the intra-species scattering length by
means of the FR depends on the atomic species: as is known, it can be done
in $^{7}$Li (see, e.g., Ref. \cite{FRlithium}), while in $^{87}$Rb loss
effects grow close to the FR point. Another possibility is offered by spinor
condensates, where the scattering lengths which determine collisions between
atoms with the same or opposite values of the hyperfine spin, $m_{F}=\pm 1$,
can be represented, respectively, as \cite{Ho} $a=a_{0}\pm a_{2}$, the
coefficients $a_{0}$ and $a_{0}$ accounting for the mean-field
(spin-independent) and spin-exchange interactions between the atoms. In this
case, the self-scattering length vanishes in the case of $a_{2}=-a_{0}$.

Besides that, the model with zero interaction inside each species
and repulsion between them may also apply to a mixture of two
ultra-cold Fermi gases \cite{Poland,Salerno}. In this connection,
it is relevant to mention that the scattering length of collisions
between fermionic atoms (such as $^{6}$Li) may also be controlled
by means of the FR \cite{FRlithium,FRfermion}. An effect of the
intra-species repulsion will be briefly considered too, with a
conclusion that it additionally stabilizes two-component GSs.

The periodic OL potential gives rise to many bandgaps in the system's
spectrum. In this work we concentrate on the most fundamental situations,
with the two components of the soliton belonging to two lowest-order gaps.
This way, we will demonstrate \textit{intra-gap} solitons, with both
components sitting in either the first or second gap, and \textit{inter-gap}
solitons, with the components belonging to the different (first and second)
gaps.

The paper is structured as follows. The model is set in Section 2. An
analytical variational approximation for 2D two-component solitons is
presented in Section 3. Direct numerical results, that identify existence
and stability regions of the intra- and inter-gap solitons, are reported in
Section 3. The stability of the 2D solitons is investigated in direct
simulations, which reveal not only stable stationary solitons, but stable
breathers too. Basic results for the 1D version of the same two-component
model are collected in Section 4; in particular, the stability of the 1D
soliton is identified via computation of eigenvalues for small perturbations
(which is more difficult in the 2D case). The paper is concluded by Section
6, where we also give estimates for actual numbers of atoms in the solitons
predicted in this work.

\section{The model}

In the mean-field approximation, the binary BEC at zero temperature is
described by a system of two coupled Gross-Pitaevskii equations (GPEs) for
the wave functions $\Psi (X,Y,Z;T)$ and $\Phi (X,Y,Z;T)$ of the two species
\cite{Pethick}:
\begin{eqnarray}
i\hbar \frac{\partial \Psi }{\partial T} &=&-\frac{\hbar ^{2}}{2m}\left(
\frac{\partial ^{2}\Psi }{\partial X^{2}}+\frac{\partial ^{2}\Psi }{\partial
Y^{2}}+\frac{\partial ^{2}\Psi }{\partial Z^{2}}\right)   \notag \\
&&+V_{0}\left[ \cos (2kX)+\cos (2kY)\right] \Psi +U(Z)\Psi +\frac{4\pi \hbar
^{2}}{m}a\left( \rho |\Psi |^{2}+|\Phi |^{2}\right) \Psi ,  \notag \\
&&  \label{GPE} \\
i\hbar \frac{\partial \Phi }{\partial T} &=&-\frac{\hbar ^{2}}{2m}\left(
\frac{\partial ^{2}\Phi }{\partial X^{2}}+\frac{\partial ^{2}\Phi }{\partial
Y^{2}}+\frac{\partial ^{2}\Psi }{\partial Z^{2}}\right)   \notag \\
&&+V_{0}\left[ \cos (2kX)+\cos (2kY)\right] \Phi +U(Z)\Phi +\frac{4\pi \hbar
^{2}}{m}a\left( \rho |\Phi |^{2}+|\Psi |^{2}\right) \Phi ,  \notag
\end{eqnarray}where $m$ is mass of of both species of atoms, $V_{0}$ and $\pi /k$ are the
amplitude and period of the OL potential, $U(Z)$ is a potential
accounting for the tight magnetic or optical confinement in the
transverse direction, that makes the condensate effectively
two-dimensional (squeezing it into a ``pancake" shape
\cite{pancake}), while $a$ and $\rho a$ (with $\rho \geq 0$) are
the scattering lengths of the inter-species and intra-species
collisions. The equations do not include an external trapping
potential in the $\left( X,Y\right) $ plane, as we are interested
in localized 2D states supported intrinsically by the interplay of
the inter-species repulsion and OL potential.

Assuming that the transverse trap gives rise to a ground-state
wave function $\chi _{0}(Z)$ with the respective energy $E_{0}$,
reduction of Eqs. (\ref{GPE}) to a normalized system of effective
two-dimensional equations follows the usual procedure, based on
averaging in the $Z$ direction and rescaling \cite{Pethick}. To
this end, we define $\left\{ kX,kY\right\} \equiv \left\{
x,y\right\} ,~\left( \hbar k^{2}/2m\right) T\equiv
t,~2mV_{0}/\left( \hbar k\right) ^{2}\equiv \varepsilon $
($\varepsilon $ measures the ratio of the height of the OL's
potential barrier to the atom's recoil energy),
and\begin{equation} \left\{ \Psi ,\Phi \right\} \equiv
e^{iE_{0}T/\hbar }\sqrt{\frac{\int_{-\infty }^{+\infty }\chi
_{0}^{2}(Z)dZ}{\int_{-\infty }^{+\infty }\chi
_{0}^{4}(Z)dZ}}\frac{k}{2\sqrt{2\pi a}}\left\{ \psi (x,y;t),\phi
(x,y;t)\right\} .  \label{PsiPhi}
\end{equation}The eventual form of the 2D equations is

\begin{eqnarray}
i\psi _{t}+\psi _{xx}+\psi _{yy}+\varepsilon \left[ \cos (2x)+\cos
(2y)\right] \psi -\left( \rho |\psi |^{2}+|\varphi |^{2}\right)
\psi &=&0,
\notag \\
&&  \label{model} \\
i\varphi _{t}+\varphi _{xx}+\varphi _{yy}+\varepsilon \left[ \cos (2x)+\cos
(2y)\right] \varphi -\left( \rho |\varphi |^{2}+|\psi |^{2}\right) \varphi
&=&0.  \notag
\end{eqnarray}

In addition to $\varepsilon $ and $\rho $, control parameters of
the normalized system are norms of both
components,\begin{equation} N_{1}=\int \int \left\vert \psi
(x,y)\right\vert ^{2}dxdy,~N_{2}=\int \int \left\vert \varphi
(x,y)\right\vert ^{2}dxdy,  \label{N12}
\end{equation}which determine the respective numbers of atoms as per Eq. (\ref{PsiPhi}):\begin{equation}
\left( N_{\mathrm{phys}}\right) _{1,2}=\frac{1}{8\pi a}\frac{\left(
\int_{-\infty }^{+\infty }\chi _{0}^{2}(Z)dZ\right) ^{2}}{\int_{-\infty
}^{+\infty }\chi _{0}^{4}(Z)dZ}N_{1,2}.  \label{numbers}
\end{equation}Norms (\ref{N12}), together with the Hamiltonian of the system, are its
dynamical invariants. As said above, we will be mostly dealing with the case
of $\rho =0$ (no intra-species interactions), hence essential parameters are
$\varepsilon $ and $N_{1,2}$ (generally, we will consider the asymmetric
case, with $N_{1}\neq N_{2}$).

Stationary solutions of Eqs. (\ref{model}) are looked for in the
usual form,\begin{equation} \psi (x,y,t)=u(x,y)\exp (-i\mu
_{1}t),~\varphi (x,y,t)=v(x,y)\exp (-i\mu _{2}t),
\label{stationary}
\end{equation}where $\mu _{1}$ and $\mu _{2}$ are real chemical potentials, and the real
functions $u(x,y)$ and $v(x,y)$ are solutions of the equations
\begin{eqnarray}
\mu _{1}u+u_{xx}+u_{yy}+\varepsilon \left[ \cos (2x)+\cos (2y)\right]
u-v^{2}u &=&0,  \notag \\
&&  \label{stationary_model} \\
\mu _{2}v+v_{xx}+v_{yy}+\varepsilon \left[ \cos (2x)+\cos (2y)\right]
v-u^{2}v &=&0.  \notag
\end{eqnarray}Linearization decouples Eqs. (\ref{stationary_model}), hence each chemical
potential must belong to one of bandgaps of the 2D linear
operator,\begin{equation} \hat{L}=\frac{\partial ^{2}}{\partial
x^{2}}+\frac{\partial ^{2}}{\partial y^{2}}+\varepsilon \left[
\cos (2x)+\cos (2y)\right]  \label{L}
\end{equation}(the spectrum of the operator is known, see, e.g., Ref. \cite{Ostrovskaya}
and Fig. \ref{spectrum_fig} below). Note that $\mu _{1}$ and $\mu _{2}$ may
be placed in different gaps, which gives rise, as said above, to
two-component \textit{inter-gap solitons}, as opposite to ones of the
\textit{intra-gap} type, with $\mu _{1}$ and $\mu _{2}$ belonging to the
same bandgap. Below, we will demonstrate that the model supports stable
solitons of both types (the consideration will actually be limited to two
lowest gaps, the intra-gap solitons being possible in both of them).

It is relevant to mention that, in nonlinear optics, one-component 1D
spatial solitons sitting in higher gaps were experimentally observed in
arrays of waveguides with the Kerr (cubic) nonlinearity \cite{Silberberg}.
Then, 2D spatial single-component lattice solitons with embedded vorticity,
belonging to the second gap, were created in a photorefractive material
equipped with a 2D photonic lattice \cite{Moti}. Moreover, in the latter
case, two-component solitons composed of a fundamental soliton in the first
gap and a vortex soliton in the second gap were also created. The most
essential difference of the photorefractive media from BEC is the saturable
character of the photorefractive nonlinearity.

\section{Variational approximation for two-dimensional solitons}

It is known that 1D and 2D solitons generated by the GPE with the
lattice potential (in the 2D case, the lattice may be either
two-dimensional or quasi-one-dimensional \cite{BBB2}) are
naturally classified, in both cases of the attractive
\cite{BBB2,BBB1} and repulsive \cite{Sakaguchi} interaction, as
\textit{tightly-bound} (TB, alias single-cell) and
\textit{loosely-bound} (LB, alias multi-cell) ones. The solitons
of the TB and LB types are essentially localized in one or several
cells of the OL potential, respectively.

TB solitons, in the 1D and 2D geometry alike (including the case of the 2D
equation with the quasi-1D lattice), may be accurately predicted by the
variational approximation (VA) \cite{BBB2,BBB1} (this approximation was
first applied to the GPE in Refs. \cite{firstVA}; a general review of the
technique can be found in Ref. \cite{Progress}). For LB solitons, the VA may
provide an adequate approximation for the soliton's central lobe only, but
not for the slowly decaying oscillatory tails \cite{OurPaper}.

We start the analysis with application of the VA to Eqs.
(\ref{stationary_model}). The Lagrangian from which the equations
are derived is $L=\int \int \mathcal{L}dxdy$, with the density
\begin{eqnarray}
\mathcal{L} &=&\mu _{1}u^{2}+\mu
_{2}v^{2}-u_{x}^{2}-u_{y}^{2}-v_{x}^{2}-v_{y}^{2}  \notag \\
&&+\varepsilon \left[ \cos (2x)+\cos (2y)\right] (u^{2}+v^{2})-u^{2}v^{2}.
\label{lagrangian_density}
\end{eqnarray}We approximate the TB solitons by a simple isotropic Gaussian \textit{ansatz},
\begin{equation}
u(x,y)=A\exp \left( -\frac{x^{2}+y^{2}}{2a^{2}}\right) ,~v(x,y)=B\exp \left(
-\frac{x^{2}+y^{2}}{2b^{2}}\right) .  \label{ansatz}
\end{equation}The substitution of the ansatz in Eq. (\ref{lagrangian_density}) and
integration yield an effective Lagrangian,
\begin{eqnarray}
L &=&\pi N_{2}\left( -\frac{1}{b^{2}}+2\varepsilon e^{-b^{2}}\right) +\pi
N_{1}\left( -\frac{1}{a^{2}}+2\varepsilon e^{-a^{2}}\right)  \notag \\
&&-\frac{N_{1}N_{2}}{\pi \left( a^{2}+b^{2}\right) }+\mu _{1}N_{1}+\mu
_{2}N_{2},  \label{lagrangian}
\end{eqnarray}where the norms of the components, $N_{1}=\pi A^{2}a^{2}$ and $N_{2}=\pi
B^{2}b^{2}$, are obtained by the substitution of the ansatz (\ref{ansatz})
in Eqs. (\ref{N12}). Below, we use, instead of $N_{1}$ and $N_{2}$, the
total and relative norms,
\begin{equation}
N\equiv N_{1}+N_{2},~N_{r}\equiv N_{1}/N_{2}  \label{NN}
\end{equation}(we define $N_{1}$ and $N_{2}$ as the smaller and larger norms,
respectively, hence $N_{r}\leq 1$).

The variational equations, $\partial L/\partial N_{1,2}=\partial L/\partial
a=\partial L/\partial b=0$, applied to Lagrangian (\ref{lagrangian}), yield
the following relations that determine parameters of the soliton within the
framework of the VA:
\begin{eqnarray}
N_{r} &=&\frac{a^{4}}{b^{4}}\frac{1-2\varepsilon b^{4}e^{-b^{2}}}
{1-2\varepsilon a^{4}e^{-a^{2}}},  \label{nr_va_eq} \\
N &=&\pi \left( 1+\frac{b^{2}}{a^{2}}\right) ^{2}(1+N_{r})\left(
2\varepsilon a^{4}e^{-a^{2}}-1\right) ,  \label{n_va_eq}
\end{eqnarray}

\begin{eqnarray}
\mu _{1} &=&-\frac{b^{2}}{a^{4}}+2\left( a^{2}+b^{2}-1\right) \varepsilon
e^{-a^{2}},  \label{mu1} \\
\mu _{2} &=&-\frac{a^{2}}{b^{4}}+2\left( a^{2}+b^{2}-1\right) \varepsilon
e^{-b^{2}}.  \label{mu2}
\end{eqnarray}Equations (\ref{nr_va_eq}) and (\ref{n_va_eq}) give rise to necessary
conditions for the existence of the soliton, $1-2\varepsilon
a^{4}e^{-a^{2}}<0$ and $1-2\varepsilon b^{4}e^{-b^{2}}<0$, from
which, in turn, it follows that that the GS does not exist unless
the OL strength $\varepsilon $ exceeds a minimum (threshold)
value, $\varepsilon _{\min }>e^{2}/8\approx 0.92$. The same
condition for the existence of GSs was predicted by the VA in the
single-component 2D model \cite{OurPaper}.

\begin{figure}[p]
\includegraphics{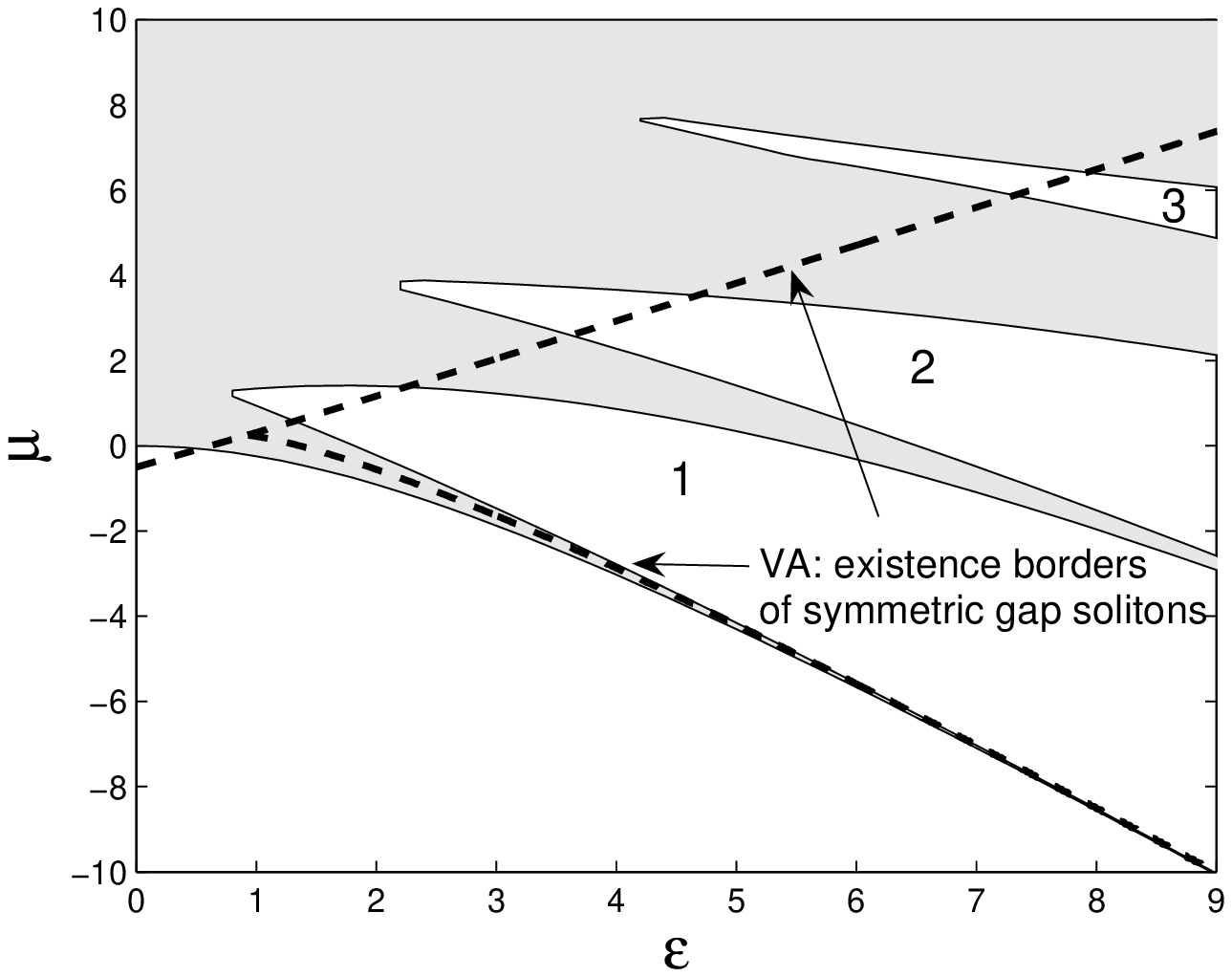}
\caption{The region of the existence of symmetric gap solitons (the area
between the two bold dashed lines), as predicted by the variational
approximation based on the Gaussian ansatz (\protect\ref{ansatz}), and the
exact bandgap structure in the two-dimensional system. Shaded and unshaded
regions are, respectively, the Bloch bands (where solitons cannot exist) and
gaps (where solitons are possible), the numbers $1$, $2$, $3$ being numbers
of the finite bandgaps. The lower unshaded region is the semi-infinite gap.}
\label{spectrum_fig}
\end{figure}

The VA does not include information about the location of bandgaps
in the system's spectrum, therefore it is necessary to check
whether the GSs predicted by the VA fall into the bandgaps. This
is shown in Fig. \ref{spectrum_fig}, which demonstrates that the
variational equations (\ref{nr_va_eq}), (\ref{n_va_eq}) and
(\ref{mu1}), (\ref{mu2}) for the symmetric GSs ($\mu _{1}=\mu
_{2}\equiv \mu $) predict the boundary of the first finite gap
(the lower bold dashed line) with surprisingly good accuracy. The
accuracy may be explained by the fact that the Gaussian ansatz
(\ref{ansatz}) provides for a good fit to the actual shape of the
GS in the first finite gap (this, in particular, means that the VA
correctly predicts the absence of solitons in the semi-infinite
gap -- the lowest one in Fig. \ref{spectrum_fig}, which extends to
$\mu \rightarrow -\infty $). On the other hand, the correlation
between the other VA-predicted soliton-existence boundary (the
upper bold dashed line in Fig. \ref{spectrum_fig}) and the exact
border of the bandgaps is very crude, due to the fact that the GSs
in higher gaps are very different from the simple ansatz
\cite{OurPaper}.

\begin{figure}[p]
\subfigure{\includegraphics[width=3in]{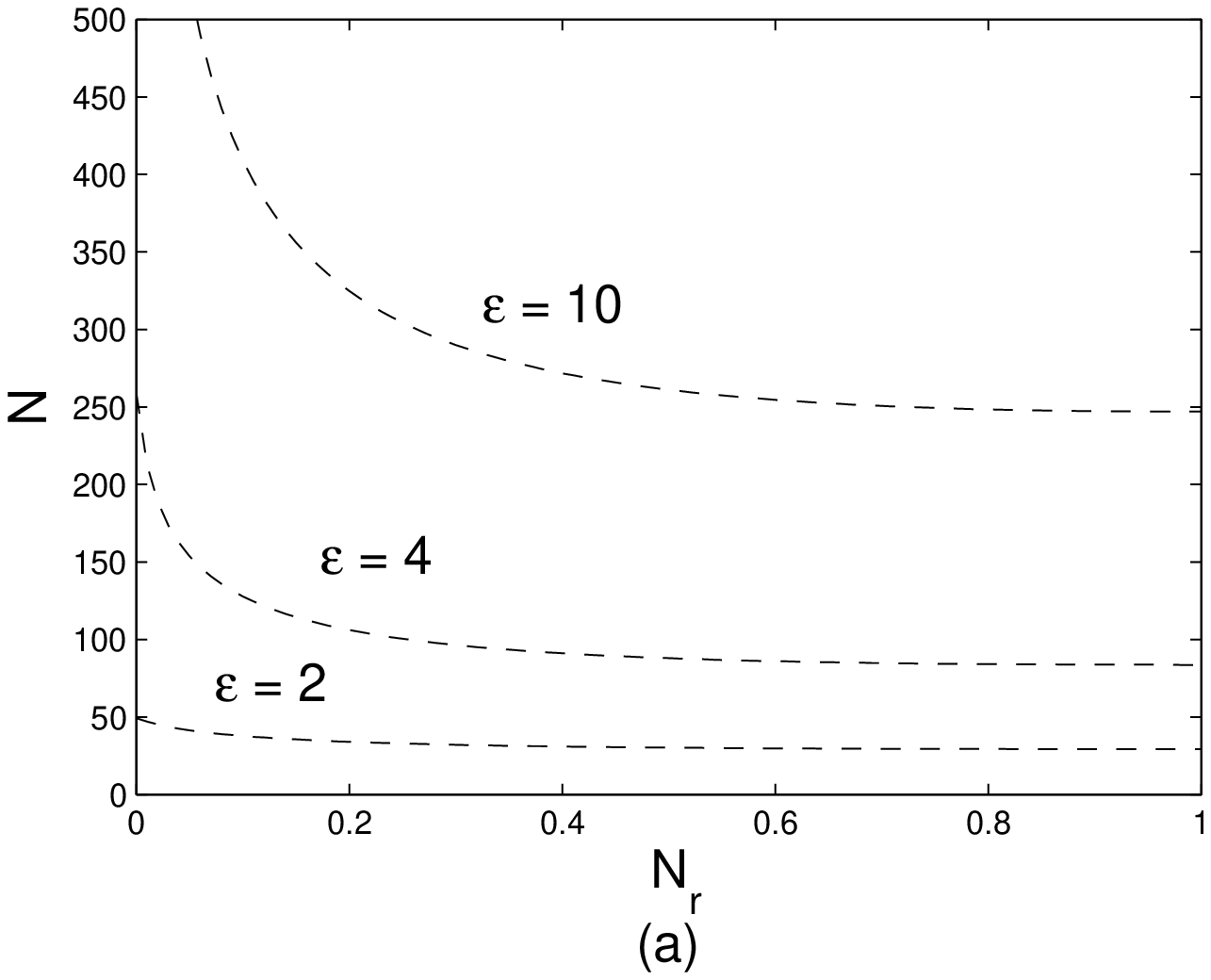}}
\subfigure{\includegraphics[width=3in]{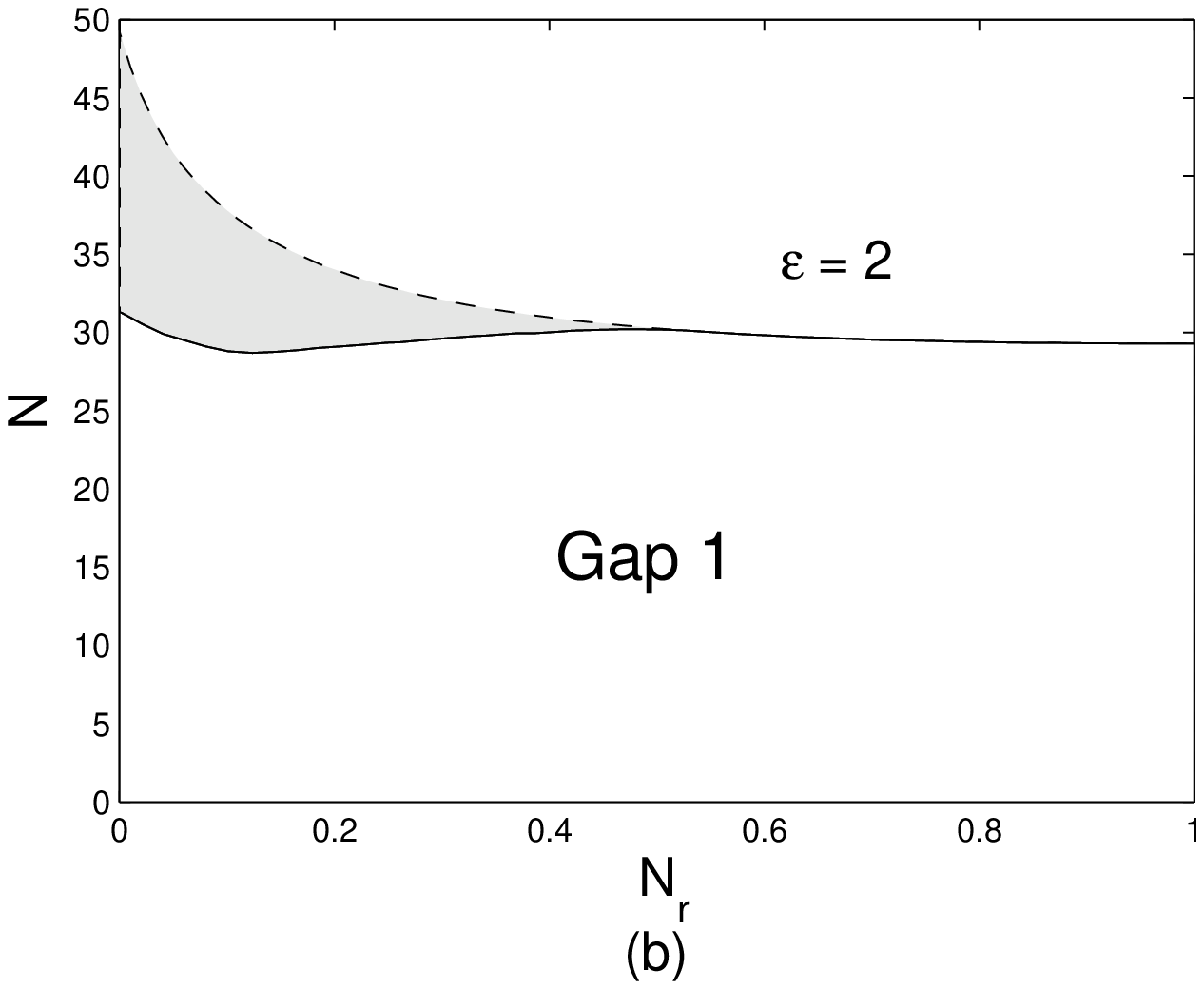}}
\newline
\subfigure{\includegraphics[width=3in]{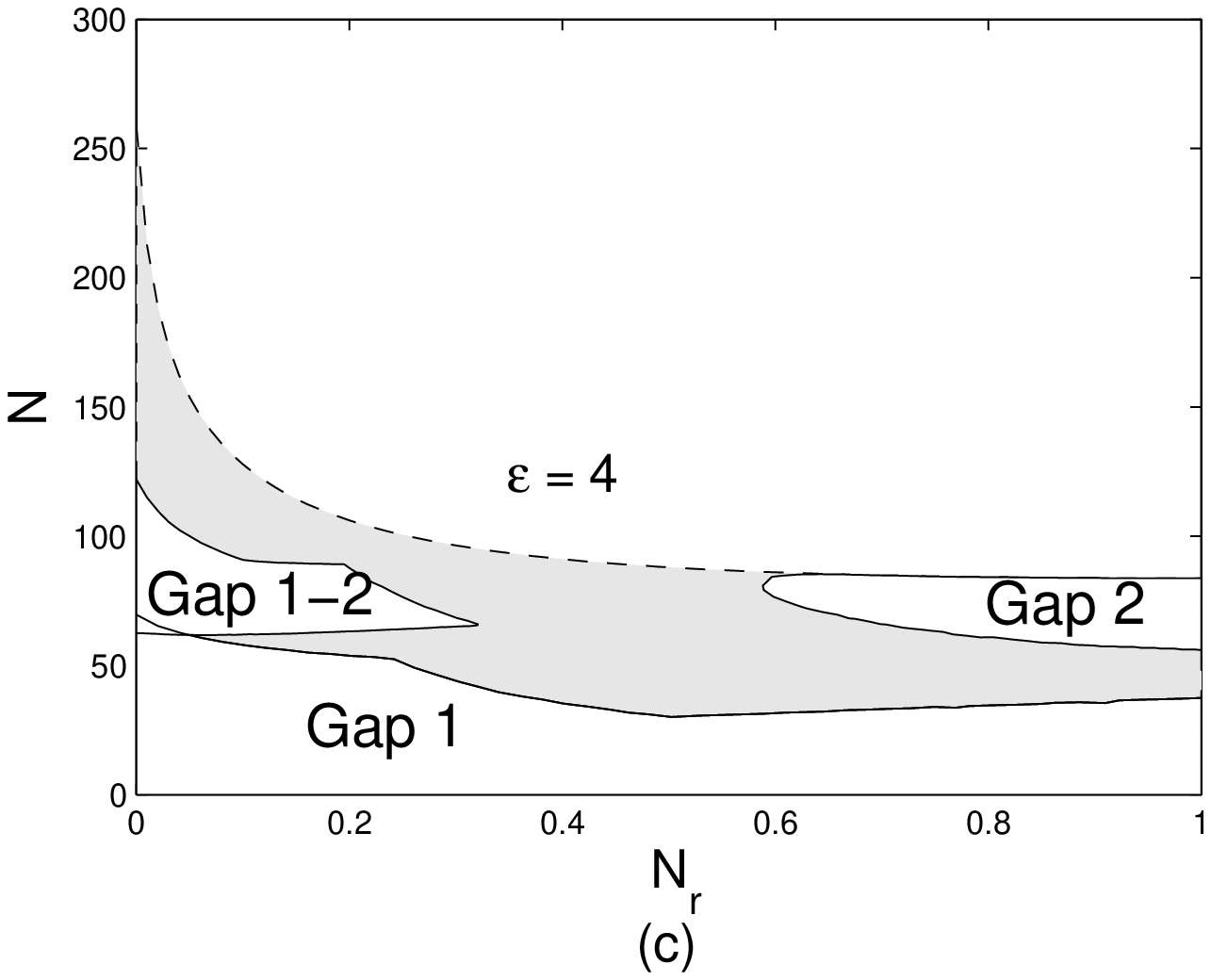}}
\subfigure{\includegraphics[width=3in]{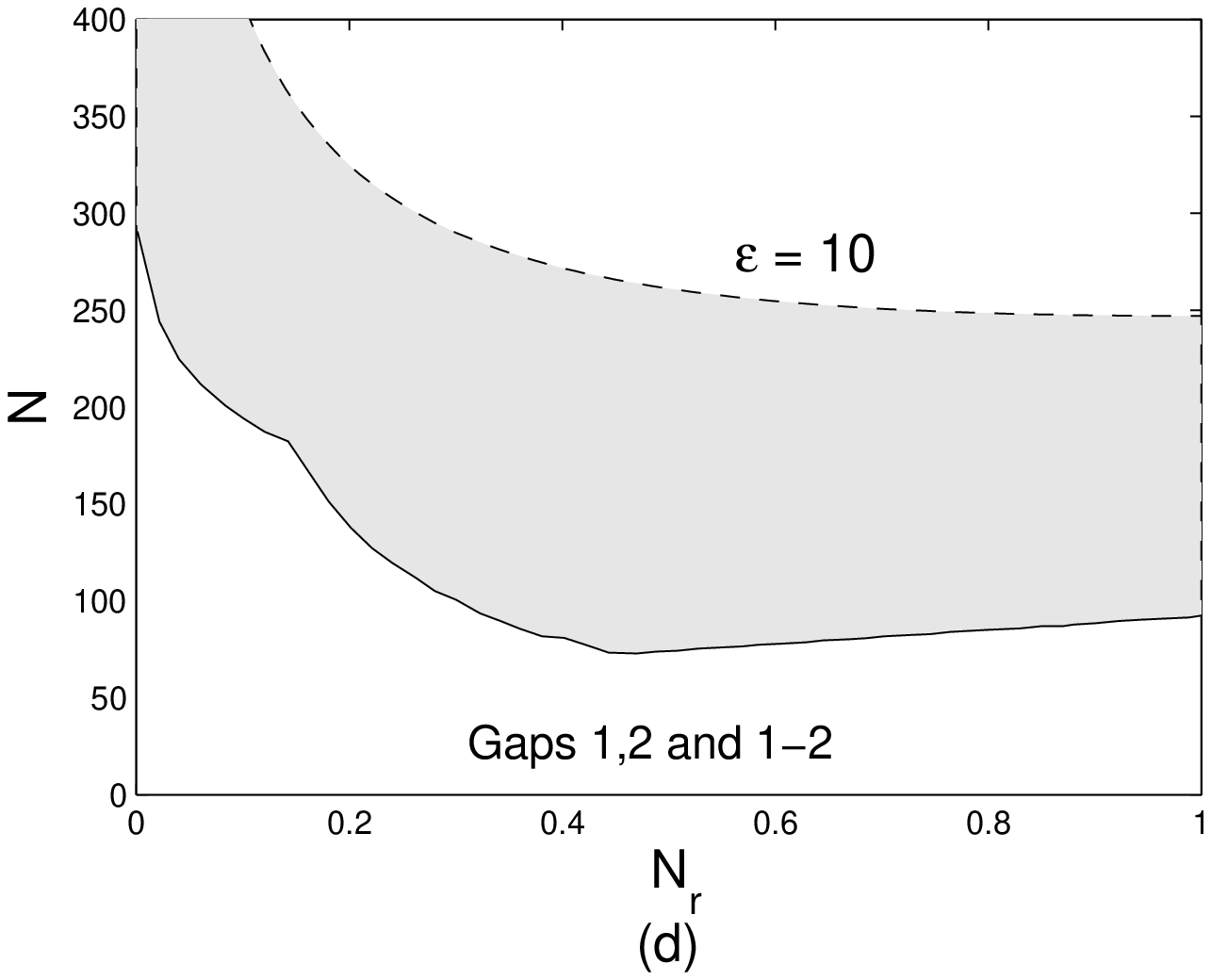}}
\caption{(a) The existence regions for the gap solitons in the
($N$,$N_{r}$) plane, as predicted (below the respective dashed
lines) by the variational approximation for different values of
the strength $\protect\varepsilon $ of the optical lattice. Panels
(b)-(d), which pertain, respectively, to $\protect\varepsilon
=2$,$~4$, and $10$, show that shaded portions of the predicted
existence regions must be excluded, as in these areas either
chemical potential, $\protect\mu _{1}$ or $\protect\mu _{2}$,
falls into a Bloch band. The remaining white portions of the
existence region in each panel are true ones, with both
$\protect\mu _{1}$ and $\protect\mu _{2}$ located in one of the
two lowest bandgaps. In panels (b) and (c), the bandgap(s) to
which the chemical potentials belong are indicated. In panel (d),
details for different gaps are not shown, because of a complex
picture of areas corresponding to each gap.} \label{existence_fig}
\end{figure}

For the description of experimentally relevant situations, Fig.
\ref{existence_fig} displays the predicted GS existence regions in
the plane of the total and relative soliton norms, $\left(
N,N_{r}\right) $ [see the definitions in Eqs. (\ref{NN})], which
includes the general asymmetric solutions with $N_{r}<1$. False
parts of the existence regions, that have either $\mu _{1}$ or
$\mu _{2}$ falling into a Bloch band (rather than into a gap), are
excluded in panels (b)-(d).

We note that, for relatively small $\varepsilon $ (however, it
must exceed the above-mentioned threshold value $\varepsilon
_{\min }$ necessary for the existence of the solitons), the VA
predicts only intra-gap solutions, with both $\mu _{1,2}$ lying in
the first finite gap. This prediction is correct, as for these
values of $\varepsilon $ the spectrum of the two-dimensional GPE
with the OL potential has only one finite gap, see Fig.
\ref{spectrum_fig}. For larger $\varepsilon $, the VA predicts
intra-gap solutions in the second gap, as well as inter-gap
solitons, with the components belonging to the different finite
bandgaps, first and second. With the further increase of
$\varepsilon $, the regions of the intra-gap solutions in the
first and second bandgaps and inter-gap solitons grow and overlap.

\section{Numerical results for two-dimensional solitons}

\subsection{Families of intra- and inter-gap solitons}

\begin{figure}[p]
\subfigure{\includegraphics[width=3in]{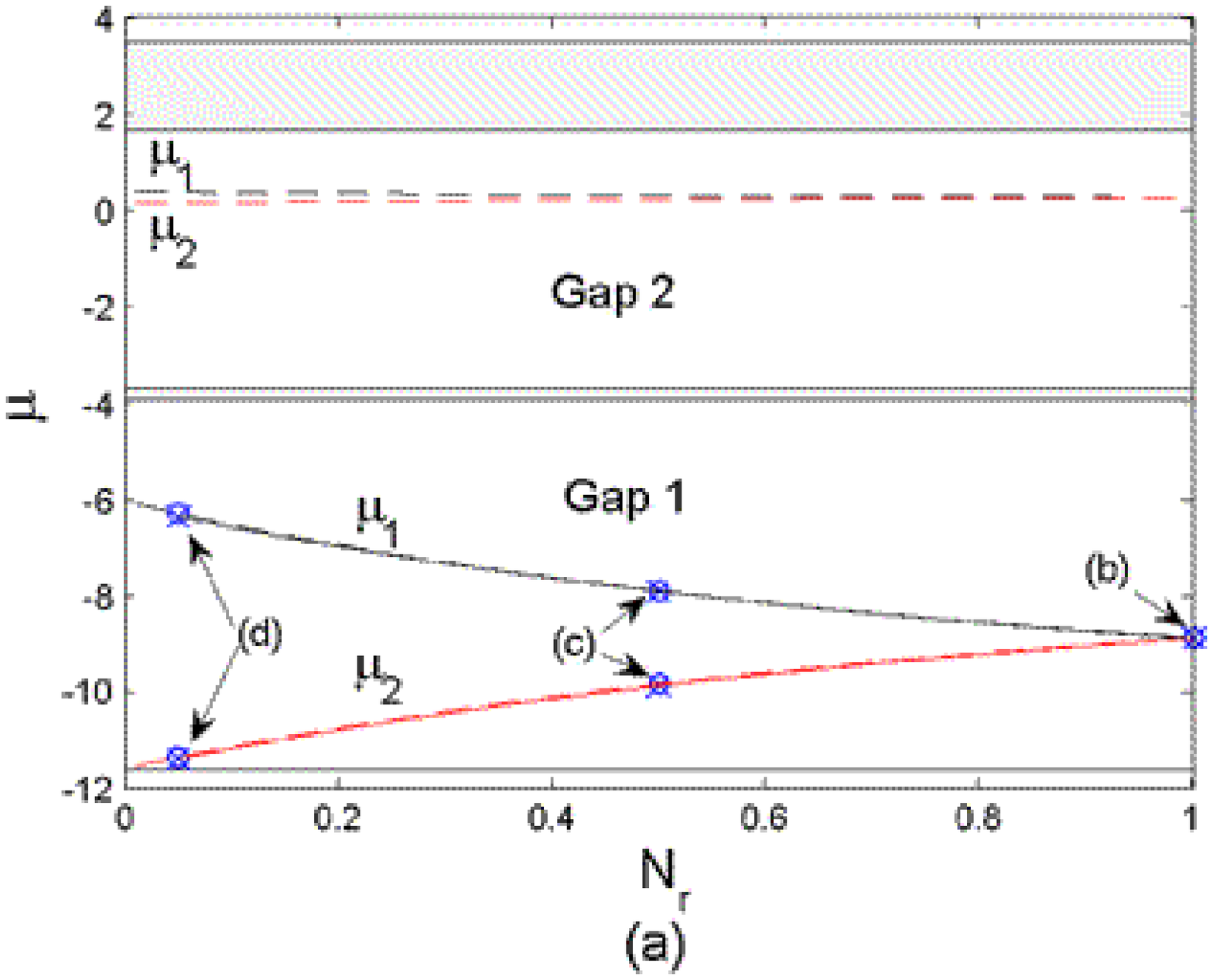}} 
\subfigure{\includegraphics[width=3in]{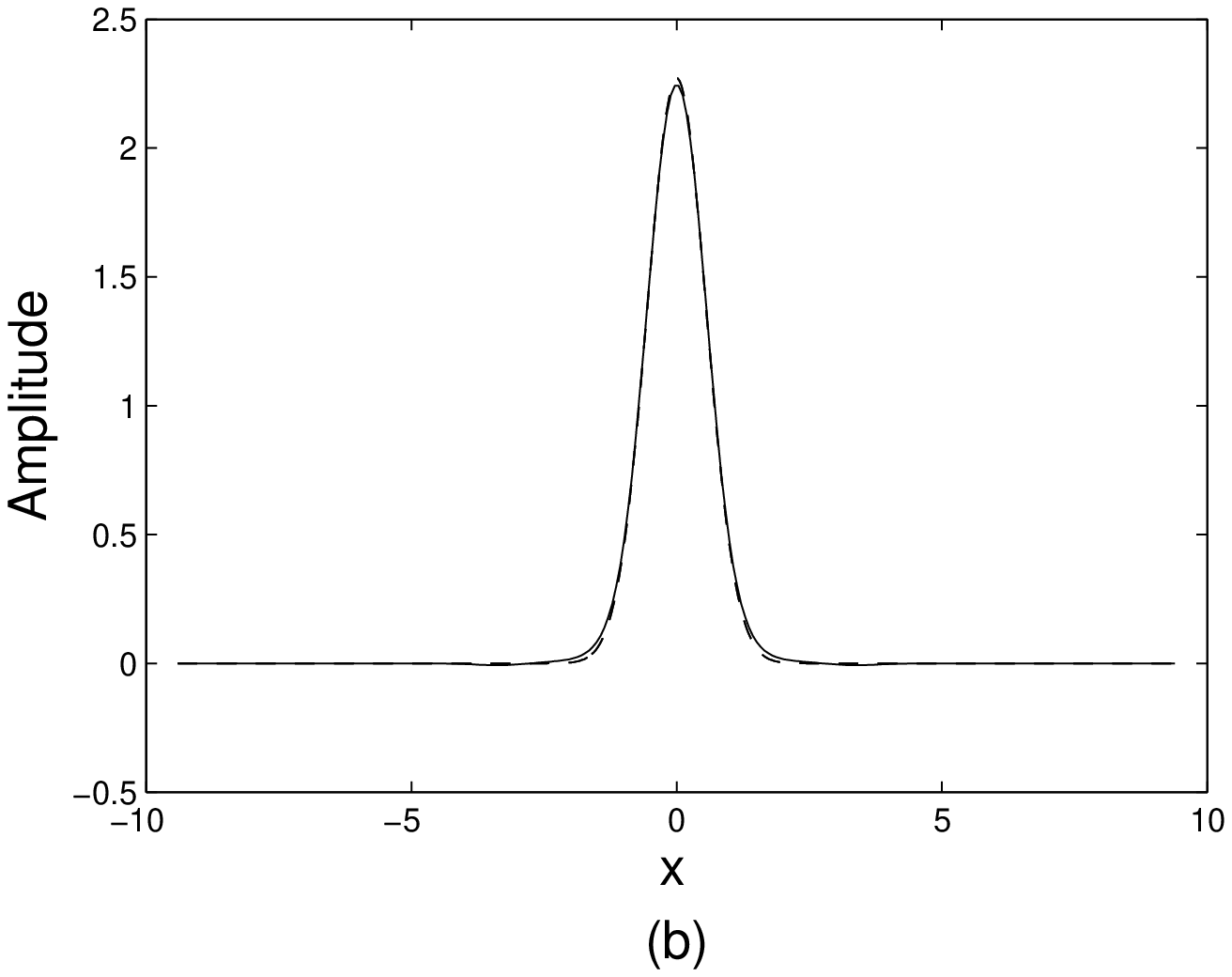}}
\subfigure{\includegraphics[width=3in]{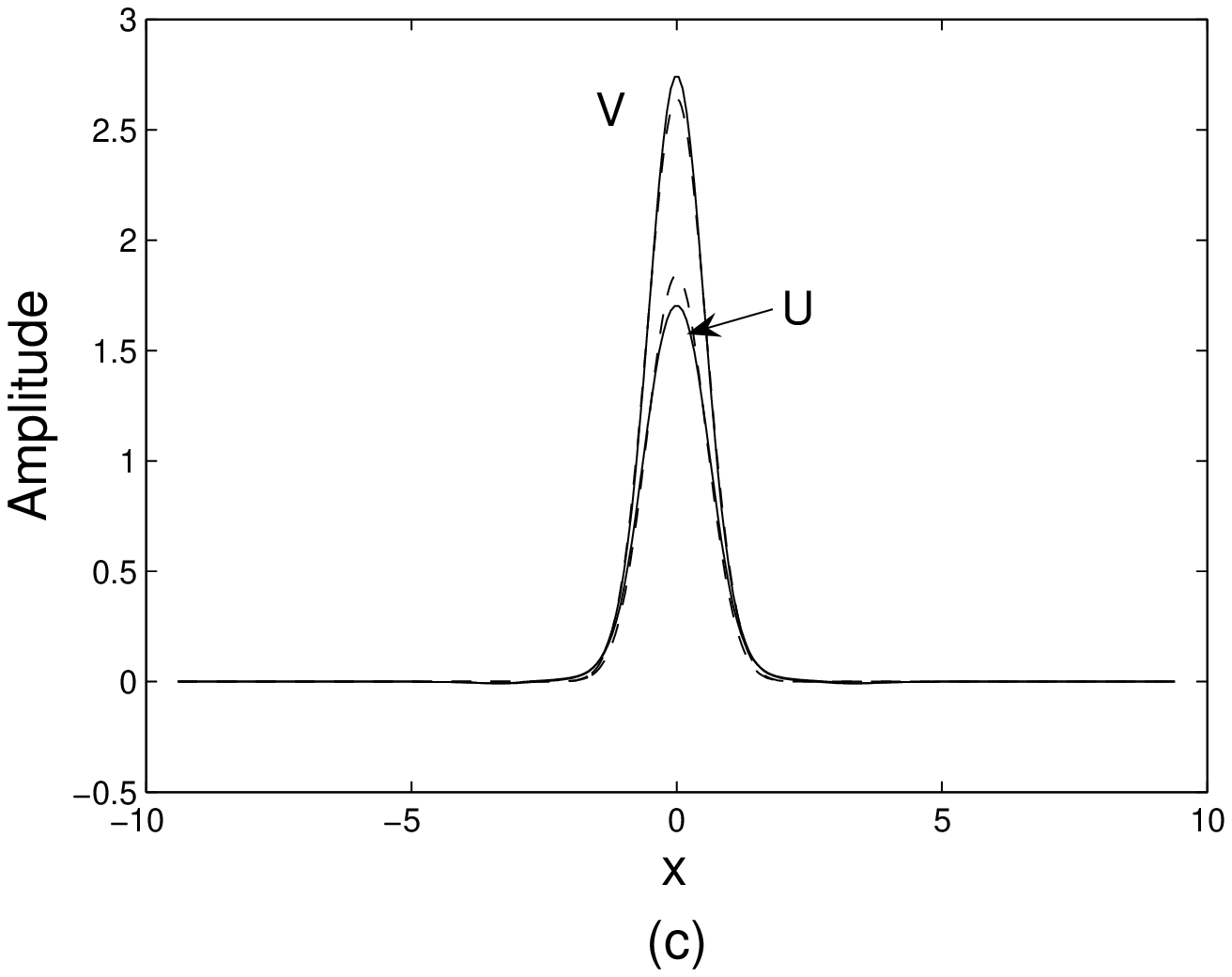}}
\centering\subfigure{\includegraphics[width=3in]{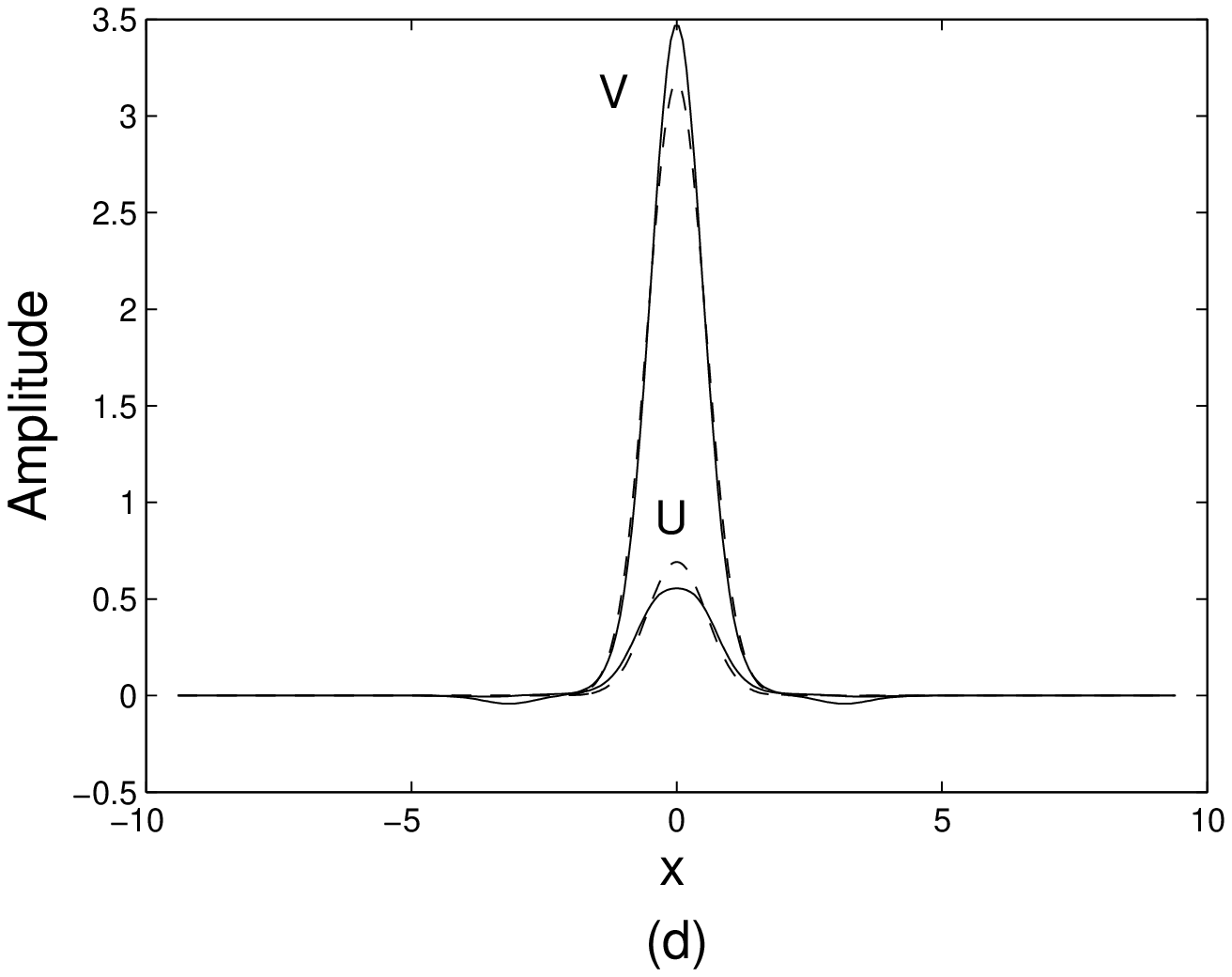}}
\caption{(Color online) Comparison of the variational
approximation and numerical solutions for two-dimensional gap
solitons in the case of $\protect\varepsilon =10$, with the total
norm $N=10$. The soliton families, as predicted by the VA and
found in the numerical form, are shown by dashed and solid lines,
respectively, in panel (a). In this panel, areas occupied by the
Bloch bands (where solitons cannot exist), are shaded. Small
circles (o) and crosses (x) designate, respectively, variational
and numerical solutions that are displayed, as examples, in the
remaining part of the figure: (b) $N_{r}=1$ (a symmetric soliton);
(c) $N_{r}=0.5$; (d) $N_{r}=0.05$. In these panels, and in
examples of the two-dimensional solitons presented below, their
cross-sections along $y=0$ are displayed. }
\label{va_vs_num_n10_ep10}
\end{figure}

Comparison with numerically found stationary solutions for the GSs
demonstrates that (as mentioned above) the applicability of the VA
is indeed limited to strongly confined TB solitons. In most cases,
the solitons with both components belonging to the first finite
gap appertain to this type, and they are predicted by the VA very
accurately, as shown in Fig. \ref{va_vs_num_n10_ep10}. Note,
however, that strongly asymmetric intra-gap solitons cannot be
found numerically in the first bandgap for values of the relative
norm smaller than $\left( N_{r}\right) _{\min }\approx 0.05$.
Nevertheless, the VA predicts asymmetric solitons for
$N_{r}<\left( N_{r}\right) _{\min }$, up to $N_{r}=0$. The latter
is an obvious artifact of the VA, as in the case of $N_{r}=0$ one
component is empty $(\psi \equiv 0 $), which makes the remaining
equation linear, hence it cannot give rise to any soliton.

On the contrary to the situation for the intra-gap solitons belonging to the
first finite bandgap, the prediction produced by the VA for the solitons
with both components sitting in the second bandgap are completely wrong: as
seen in the top panel of Fig. \ref{va_vs_num_n10_ep10}, the formally
predicted family of the intra-gap solitons in the second gap has no
numerically found counterpart.

\begin{figure}[p]
\subfigure{\includegraphics[width=3in]{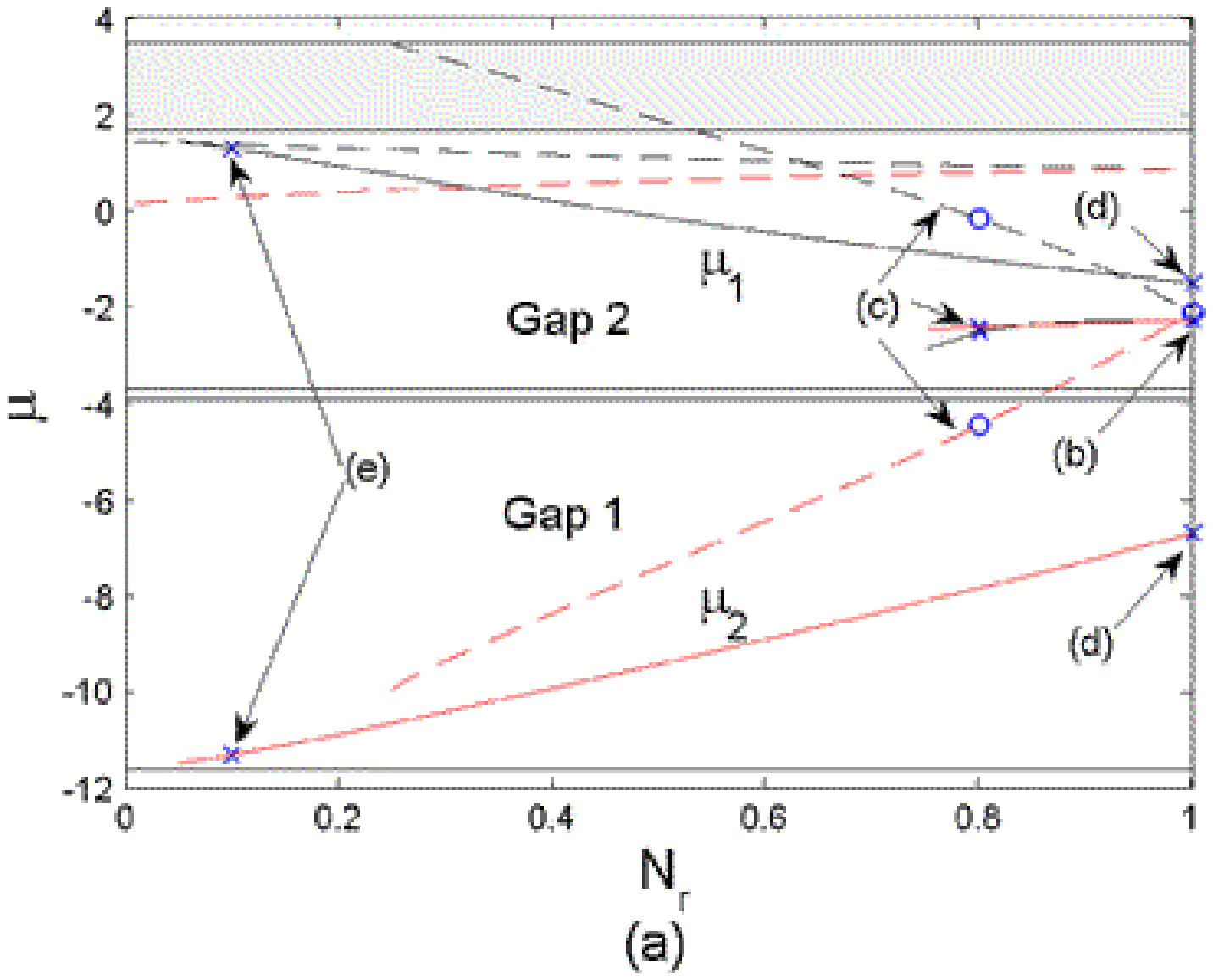}} 
\subfigure{\includegraphics[width=3in]{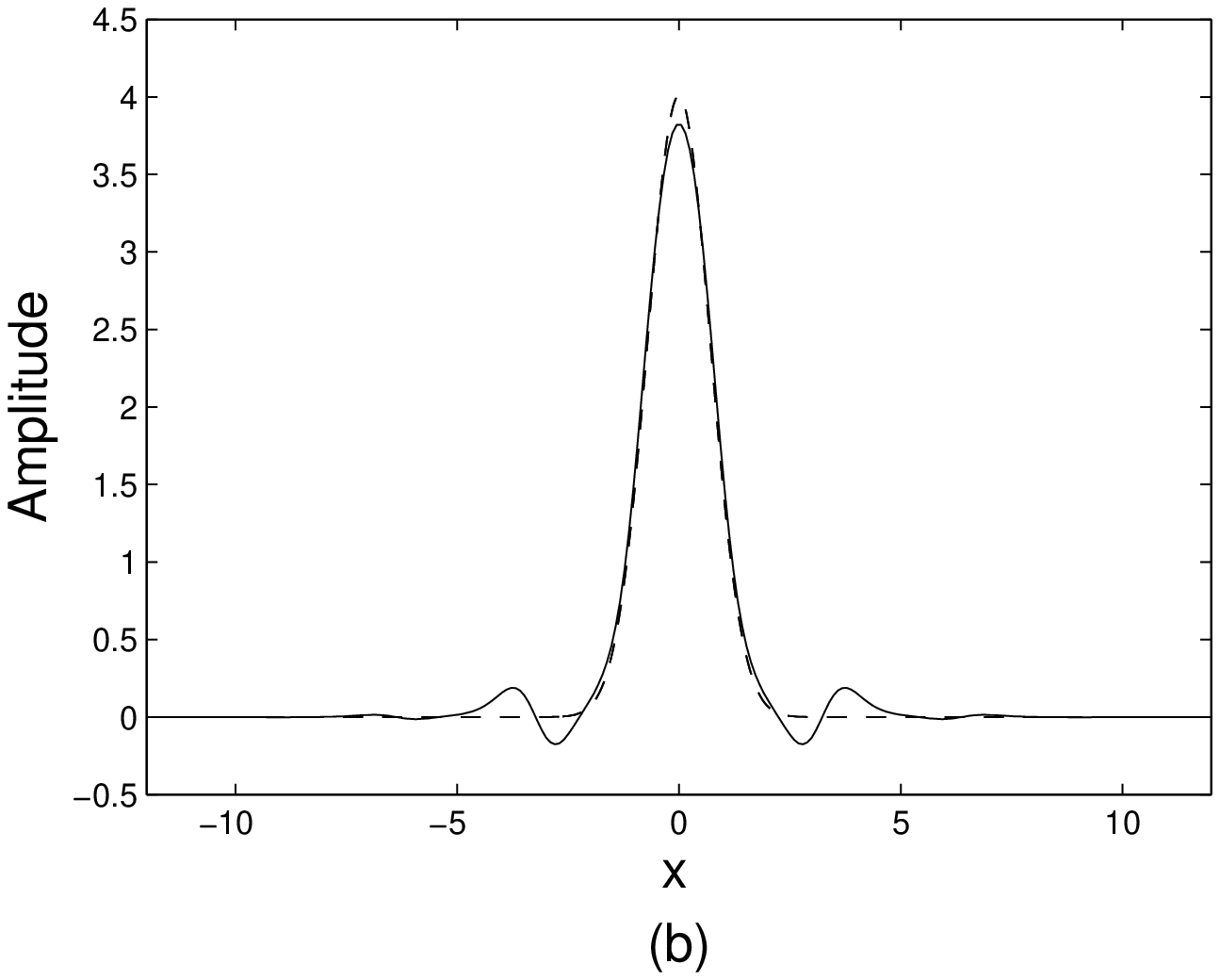}}
\subfigure{\includegraphics[width=3in]{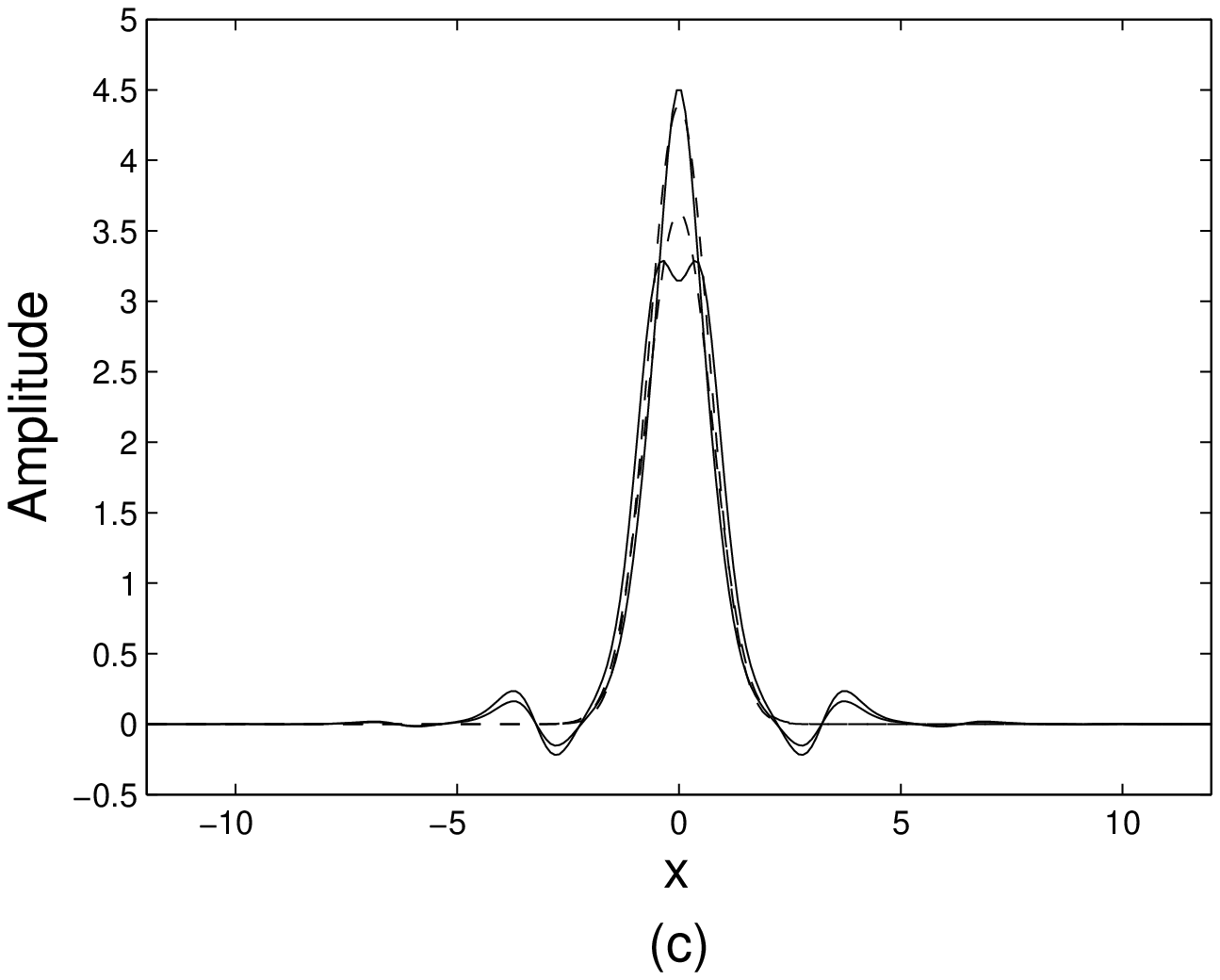}}
\subfigure{\includegraphics[width=3in]{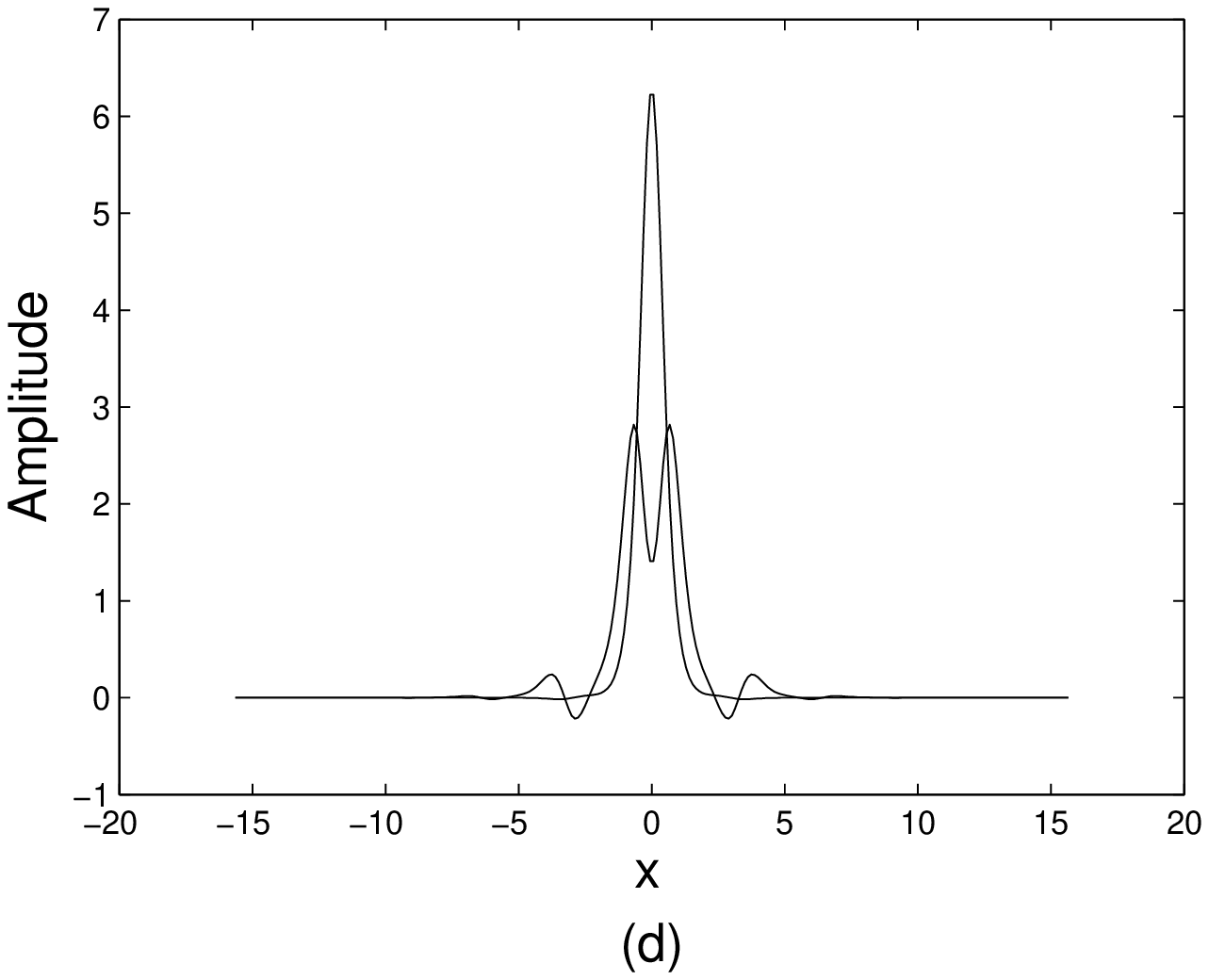}}
\subfigure{\includegraphics[width=3in]{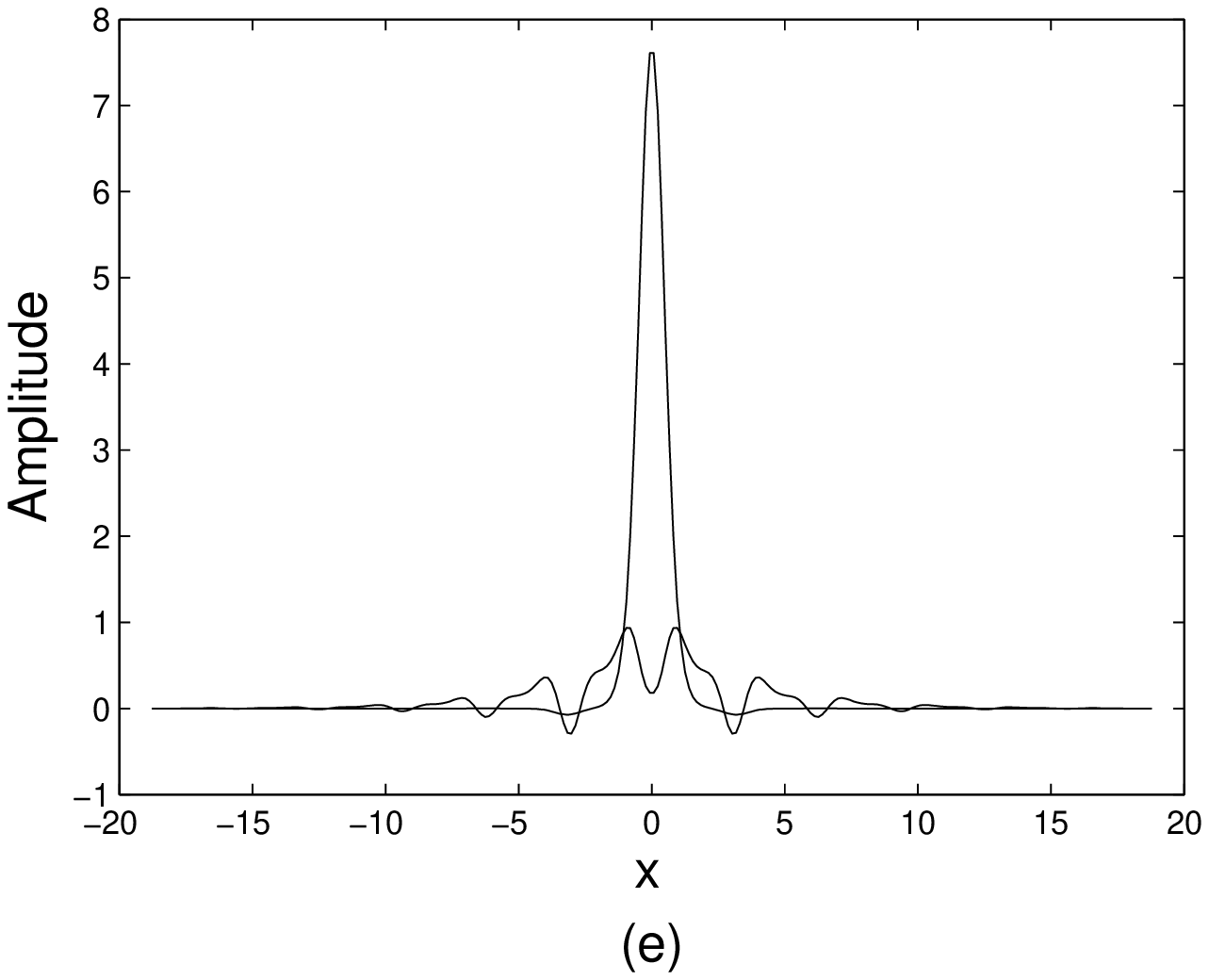}}
\caption{(Color online) The same as in the previous figure, but
for $N=50$. Panels (b) and (c): examples of the intra-gap
solitons, for $N_{r}=1$ and $N_{r}=0.8$, respectively, which
belong to the second bandgap. Panels (d) and (e): examples of
inter-gap solitons, with $N_{r}=1$ and $N_{r}=0.1$, respectively
[in both these cases, the loosely-bound component belongs to the
second (upper) bandgap, and in the latter case ($N_{r}=0.1$) the
larger norm is in the first (lower) bandgap]. Note that the
inter-gap soliton with $N_{r}=1$ is \emph{not} a symmetric one,
even if the norms of its two components are equal.}
\label{va_vs_num_n50_ep10}
\end{figure}

Increase of the total norm $N$ pushes the solution to the higher
(second) bandgap. Figure \ref{va_vs_num_n50_ep10} compares the
numerical results and VA for such a situation (with $N$ five times
as large as in Fig. \ref{va_vs_num_n10_ep10}). We observe that,
while the VA is good for the symmetric solitons ($N_{r}=1$), it
fails for more general asymmetric solutions -- both intra-gap
solitons belonging to the second gap, and inter-gap ones.

Figure \ref{va_vs_num_n50_ep10} also shows that, as the asymmetry
coefficient $N_{r}$ gets smaller, the soliton's component with the
smaller norm loses its single-peaked shape, see panels (c)-(e). We
stress that no intra-gap soliton belonging to the second gap could
be found (numerically) for $N_{r}<0.75$. The latter is seen in
Fig. \ref{va_vs_num_n50_ep10}, where the branches corresponding to
the intra-gap solution family terminate at $N_{r}\approx 0.75$.

\begin{figure}[p]
\includegraphics{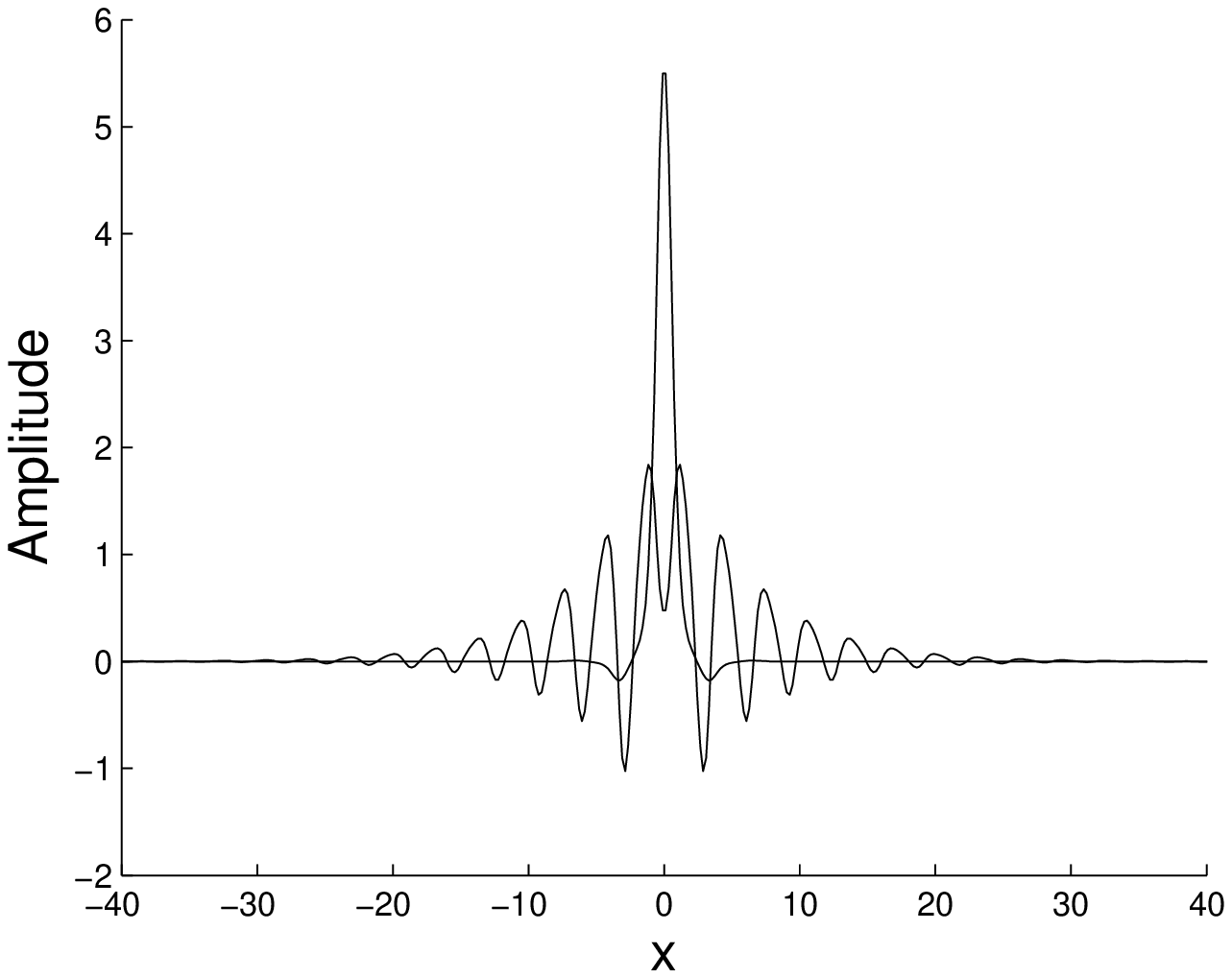}
\caption{An example of an inter-gap soliton, built as a bound
state of tightly- and loosely bound components with equal norms
($N_{r}=1$), for $\protect\varepsilon =4$ and $N=70$. The
corresponding chemical potentials are $\protect\mu _{1}=$ $-1.64$
and $\protect\mu _{2}=$ $3.3$.} \label{num_ep4}
\end{figure}

The findings presented in Fig. \ref{va_vs_num_n50_ep10} include inter-gap
solitons [in panels (d) and (e)], with $\mu _{2}$ in the first gap and $\mu
_{1}$ in the second. As seen from panel (a) in the figure, these solitons
could not be predicted by the VA. This is explained by the fact that at
least one component of the inter-gap soliton has a loosely-bound shape,
hence the Gaussian ansatz (\ref{ansatz}) is irrelevant for it. A typical
example of the inter-gap soliton built as a TB-LB bound pair is shown in
Fig. \ref{num_ep4}, for a still larger total norm, $N=70$. As well as in the
above examples, the TB and LB\ components reside in the lower and upper
gaps, respectively.

\begin{figure}[p]
\subfigure{\includegraphics[width=3in]{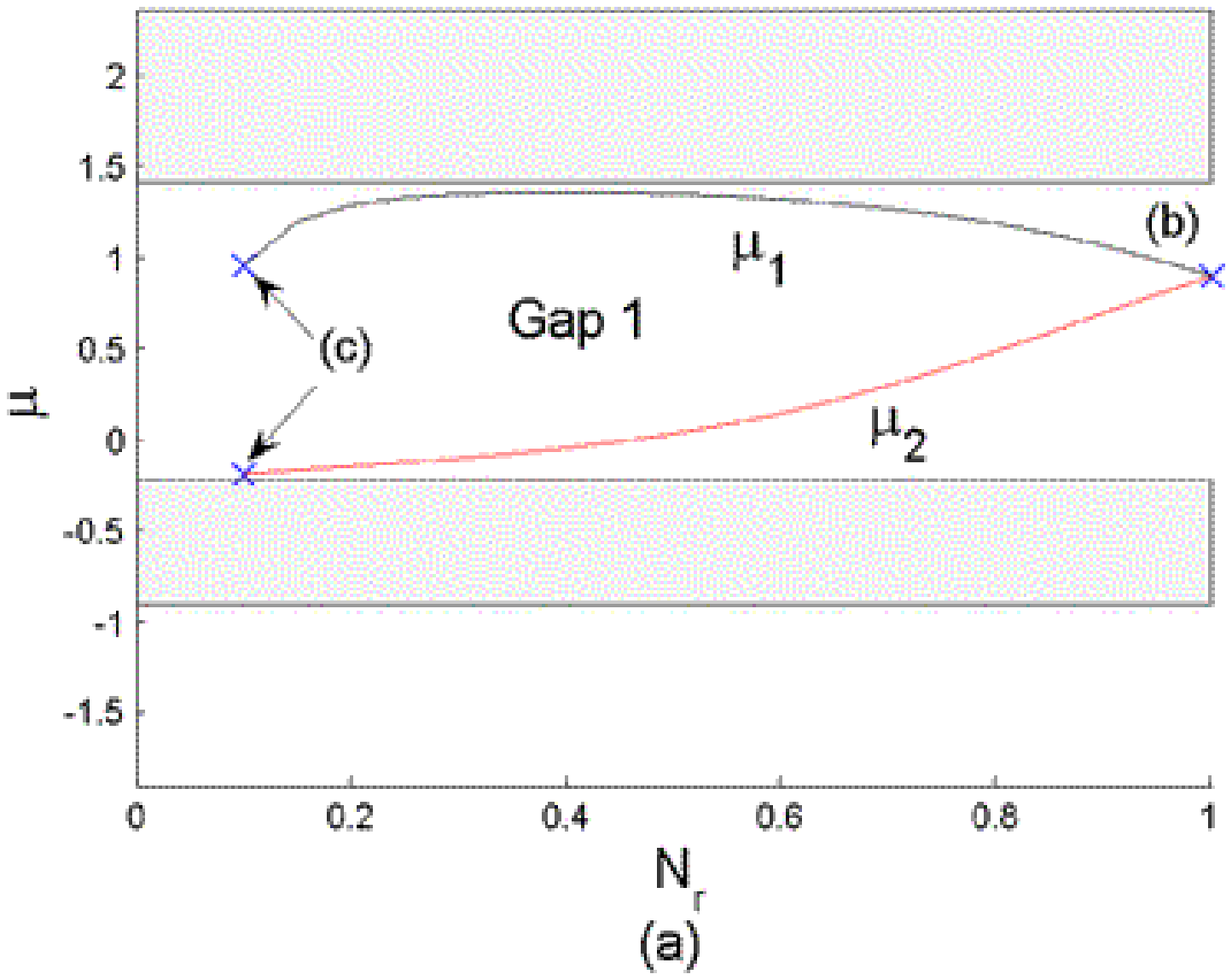}} 
\subfigure{\includegraphics[width=3in]{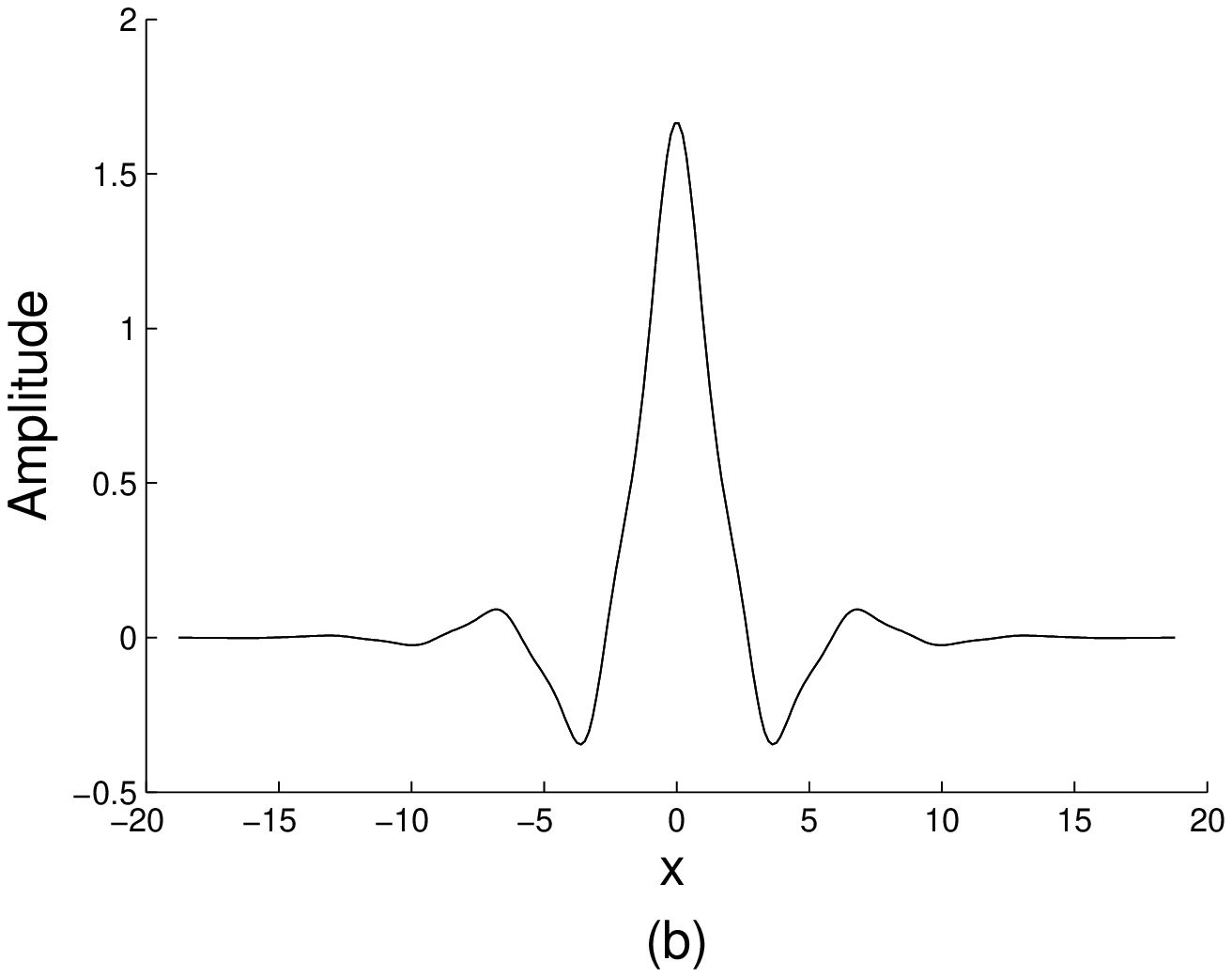}}
\subfigure{\includegraphics[width=3in]{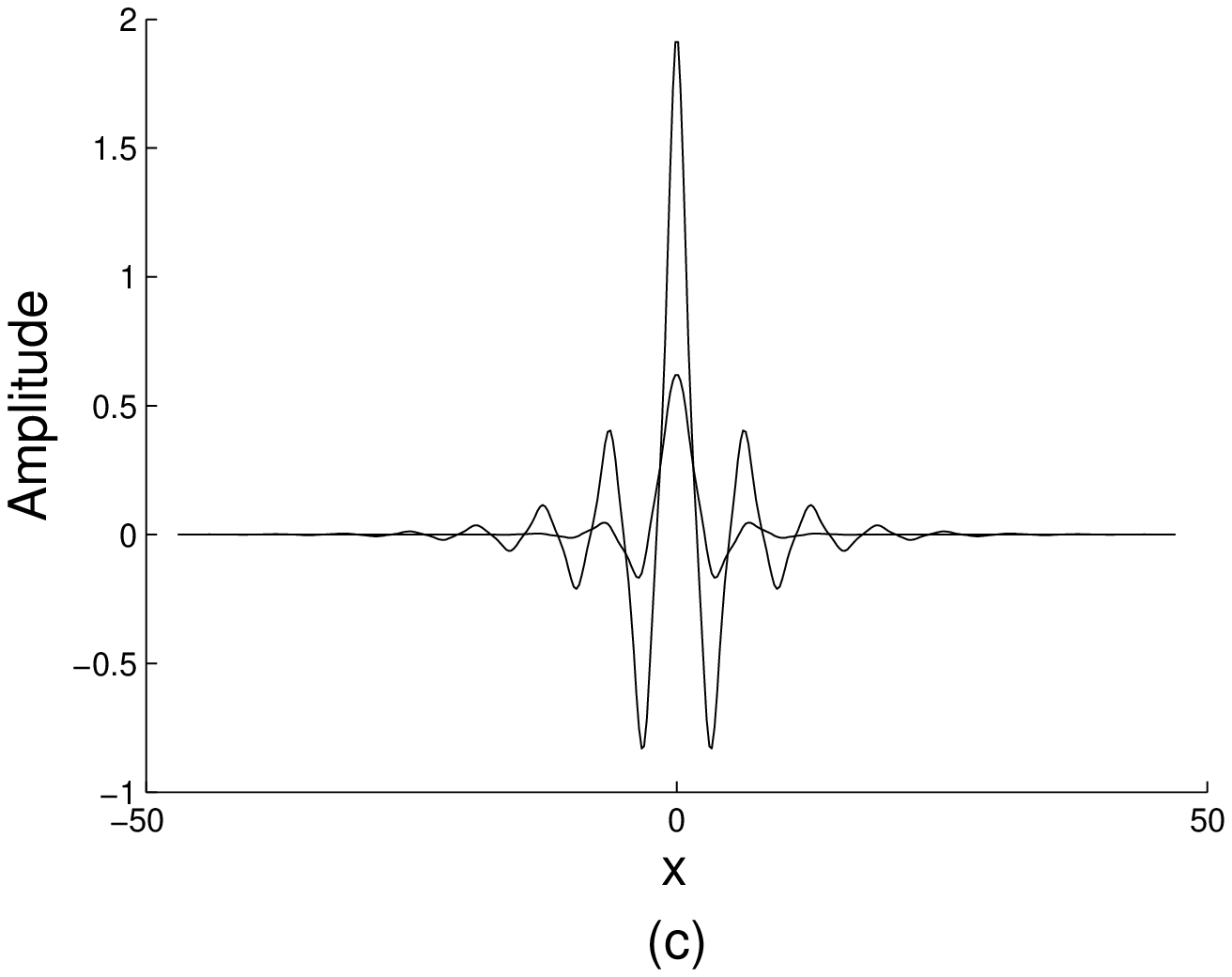}}
\caption{(Color online) Stable intra-gap solitons numerically
found for $\protect\varepsilon =2$ (in this case, the system's
spectrum contains only the first finite bandgap), with the total
norm $N=20$. (a) The family of the gap-soliton solutions. (b) A
tightly-bound symmetric soliton with $N_{r}=1$. (c) An example of
a strongly asymmetric soliton, with $N_{r}=0.1$, in the form a
bound state of loosely- and tightly bound components, with the
larger norm sitting in the \emph{loosely-bound} component.}
\label{num_ep2}
\end{figure}

Intra-gap solitons may also be TB-LB bound states. Figure \ref{num_ep2}
demonstrates how a symmetric TB soliton (with $N_{r}=1$) transforms into a
TB-LB pair with the decrease of $N_{r}$, this time (on the contrary to the
above examples of the inter-gap TB-LB bound states) the LB component having
a \emph{larger} norm, which is, in fact, a rule for asymmetric intra-gap
solitons, as evidenced by our numerical results obtained with other values
of the parameters.

With even larger norms (for instance, $N=150$), intra-gap solitons in the
form of LB-LB bound pairs with very slowly decaying tails were found too,
for $N_{r}>0.85$, i.e., they are nearly symmetric states. However, such
solitons are unstable.

\subsection{Global existence and stability diagrams for the solitons}

\begin{figure}[p]
\subfigure{\includegraphics{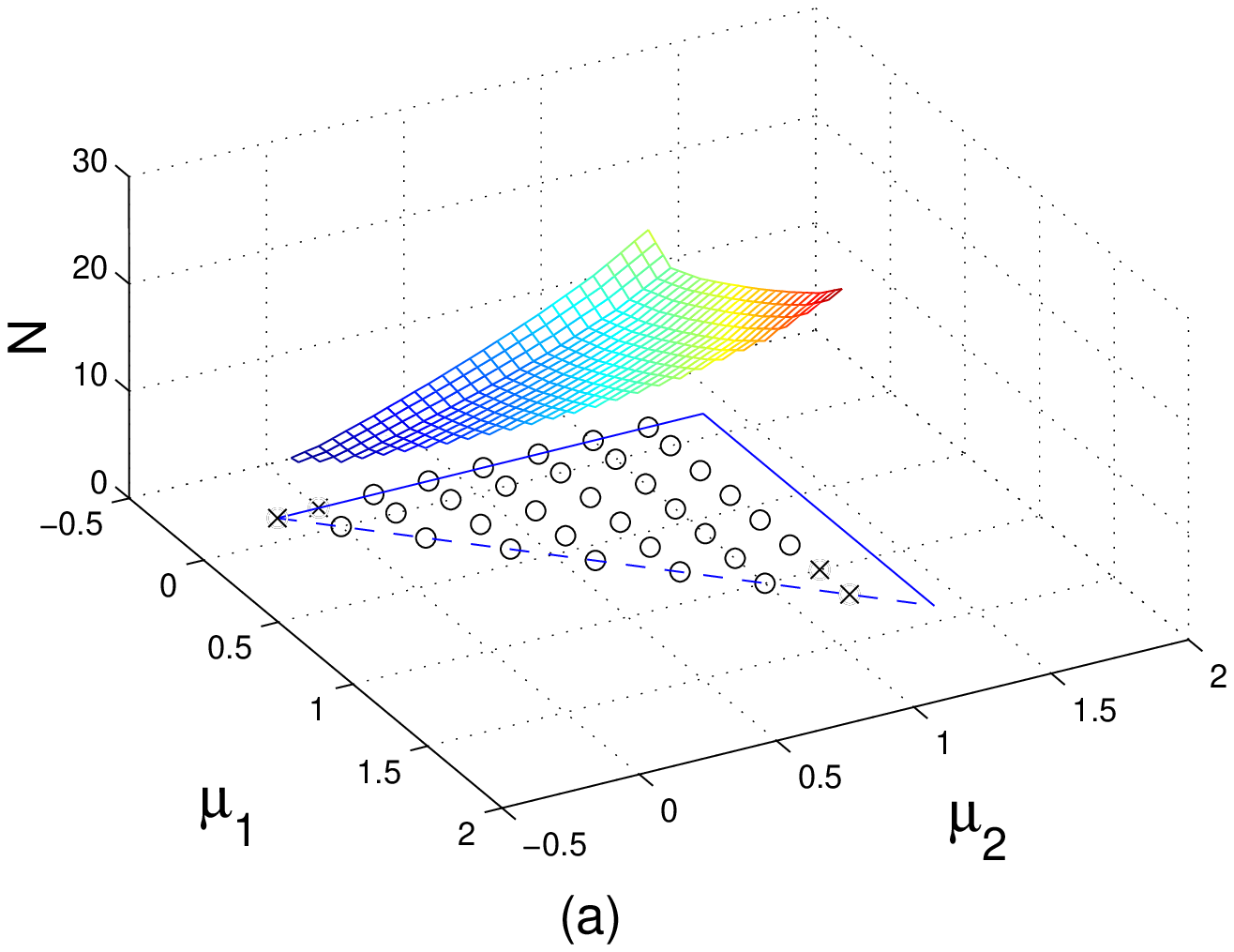}} \newline
\subfigure{\includegraphics{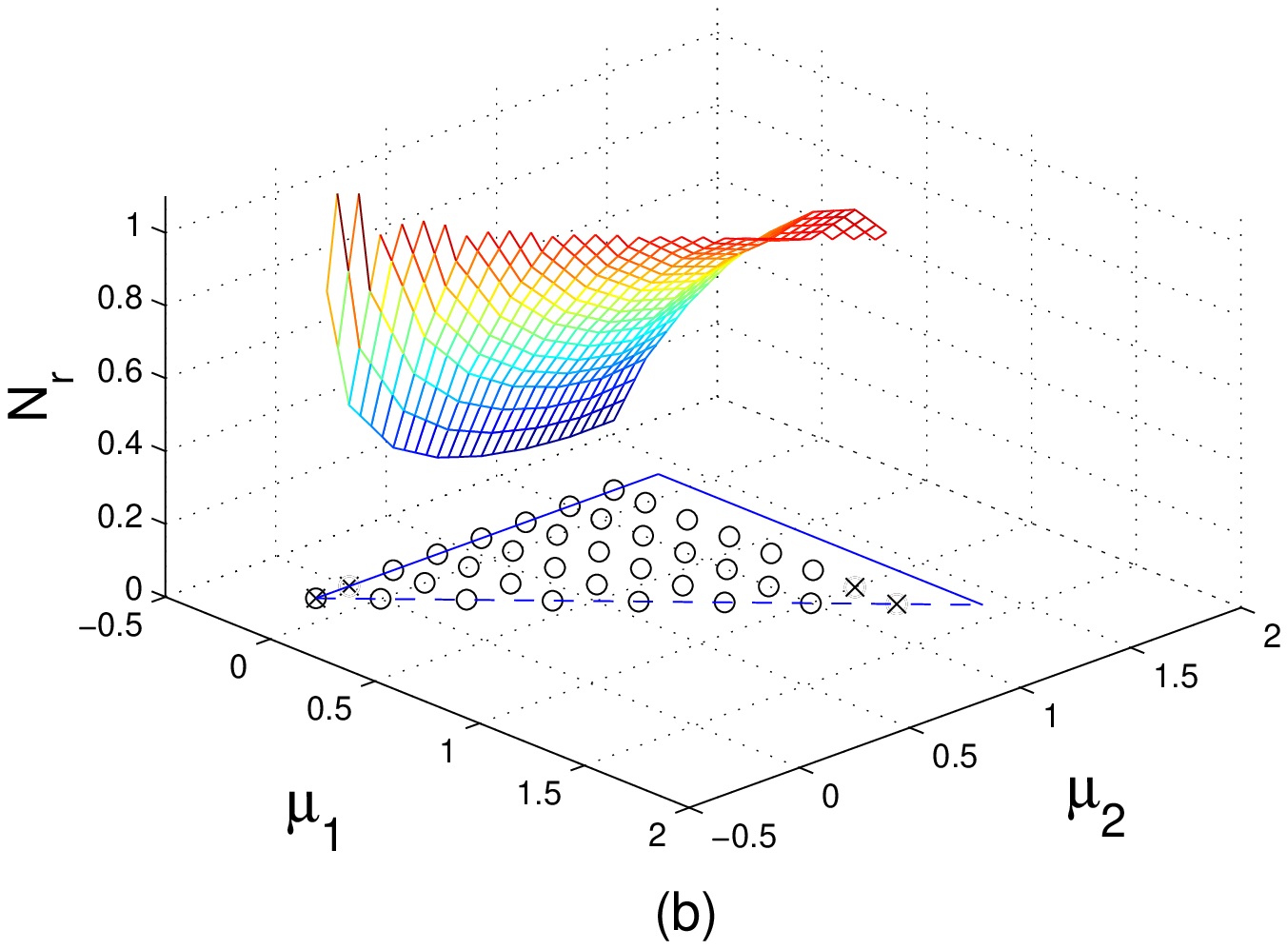}}
\caption{(Color online) The global existence and stability diagram
for the two-component intra-gap solitons in the weak
two-dimensional lattice potential, with $\protect\varepsilon =2$
(the spectrum of the 2D Gross-Pitaevskii equation contains only
one finite bandgap in this case). The total norm of the soliton,
$N=N_{1}+N_{2}$, (a), and the relative norm, $N_{r}=N_{1}/N_{2}$,
(b), are shown vs. the chemical potentials of the two components,
$\protect\mu _{1}$ and $\protect\mu _{2}$. Solid lines in the
$\left( \protect\mu _{1},\protect\mu _{2}\right) $ plane are the
bandgap's borders, and the dashed diagonal is the symmetry axis,
$\protect\mu _{1}=\protect\mu _{2}$ (the identical mirror-image
region on the other side of the diagonal is not shown). Stable and
unstable solitons are designated, respectively, by small circles
(o) and crosses (x).} \label{graph_3d_ep2}
\end{figure}

\begin{figure}[p]
\subfigure{\includegraphics{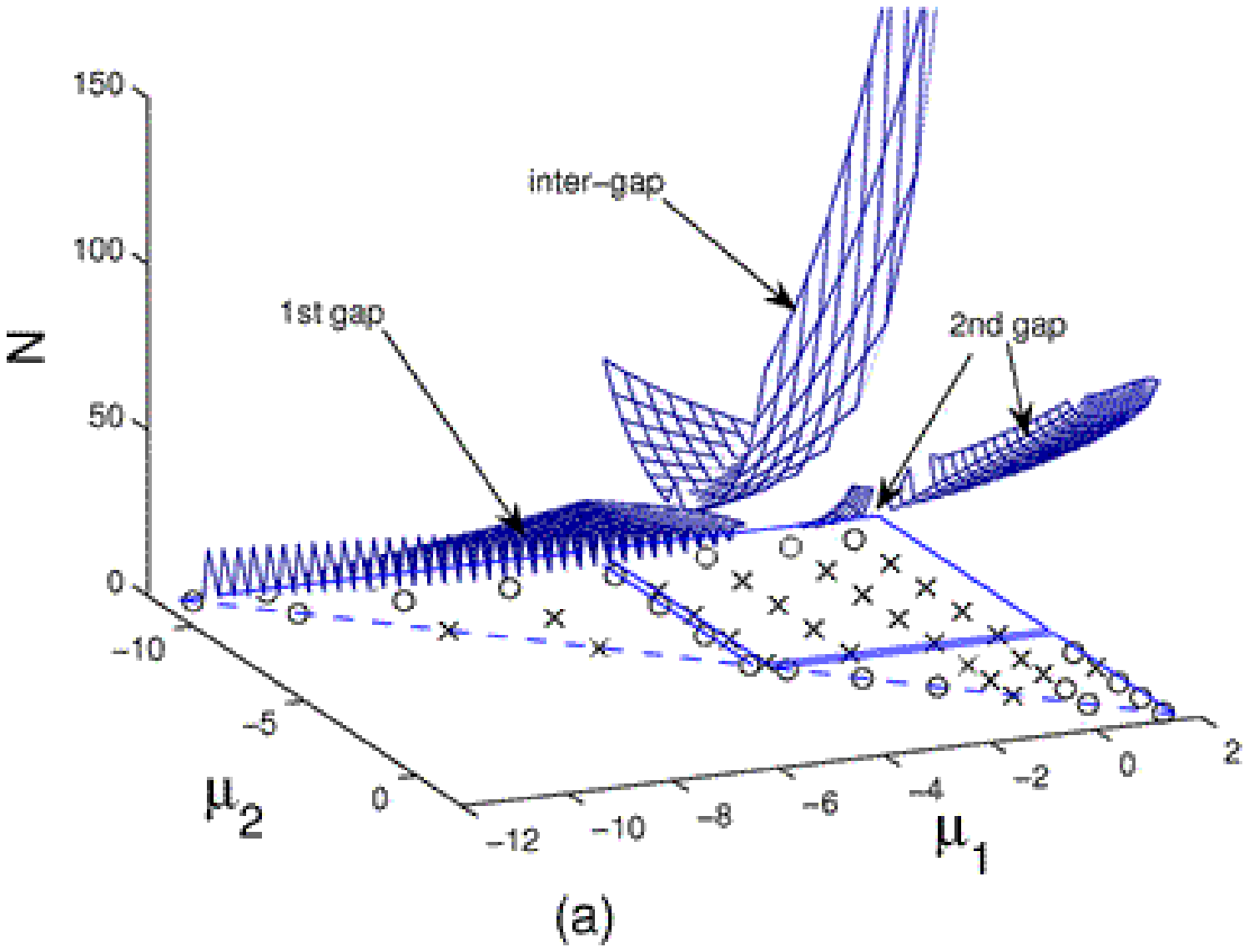}} \newline
\subfigure{\includegraphics{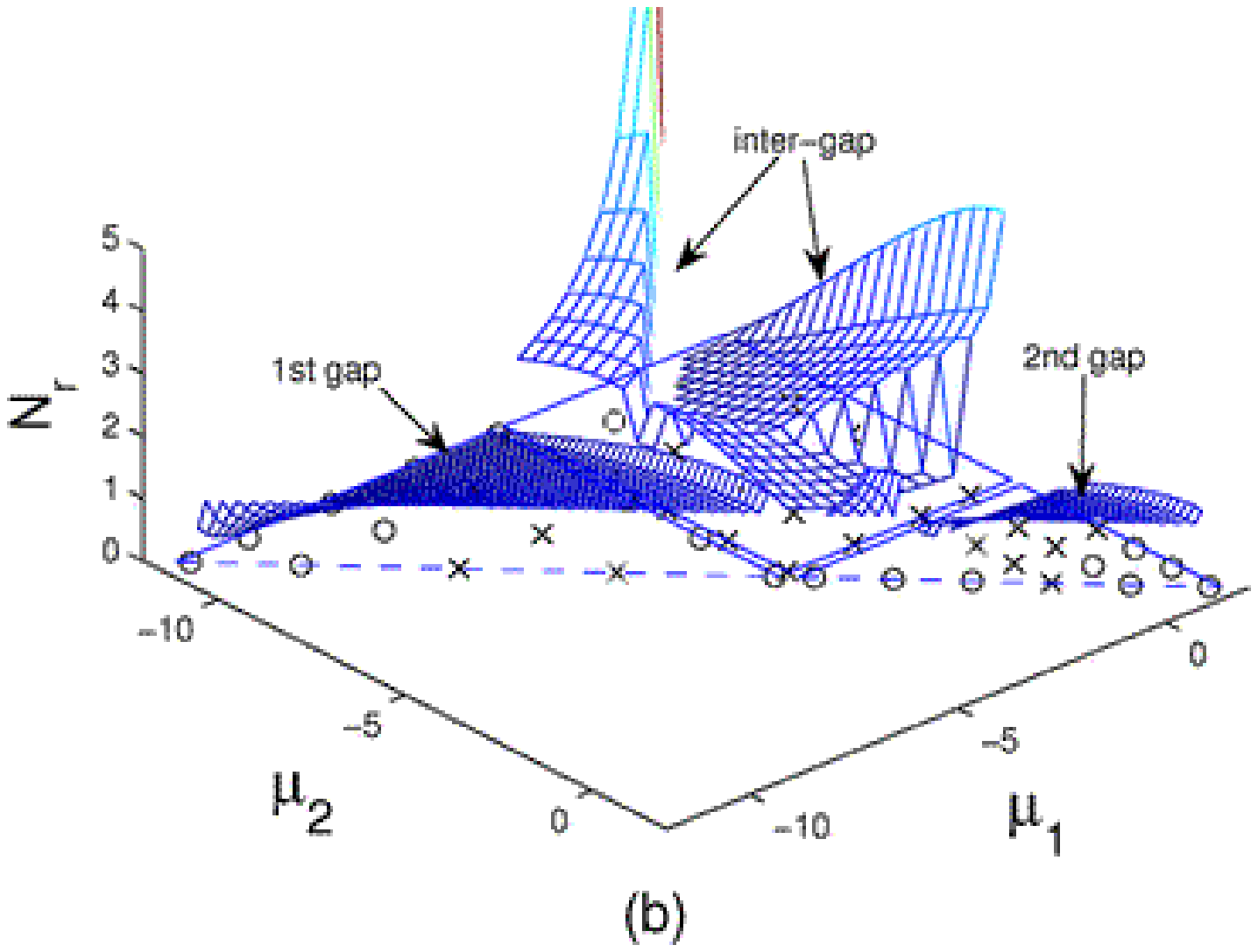}}
\caption{(Color online) The existence and stability diagram for
the two-component gap solitons in a relatively strong lattice,
with $\protect\varepsilon =10$ (in this case, the 2D
Gross-Pitaevskii equation gives rise to two finite bandgaps,
therefore the diagram includes the family of inter-gap solitons).
The notation is the same as in the previous figure, with labels
additionally indicating the type of the solitons on each disjoint
solution surface (labels ``1st gap" and ``2nd gap" pertain to
intra-gap solitons belonging to the respective gaps). Double solid
lines separating the bandgaps represent narrow Bloch bands between
the gaps. } \label{graph_3d_ep10}
\end{figure}

Results of systematic investigation of the existence and stability
of the two-component two-dimensional GSs are summarized in Fig.
\ref{graph_3d_ep2}, which is a typical example for the weak OL,
with $\varepsilon =2$, that supports only one finite bandgap in
the spectrum of the 2D model, and in Fig. \ref{graph_3d_ep10},
which represents relatively strong OLs (it has $\varepsilon =10$,
which admits two distinct finite bandgaps).

Figure \ref{graph_3d_ep2} shows that the entire gap is populated
with solitons which are stable, except when both chemical
potentials $\mu _{1}$ and $\mu _{2}$ are close to either the lower
or upper bandgap edge. The stability was verified by direct
simulations of the underlying GPEs (\ref{model}), with an initial
perturbation imposed on the soliton by dislocating centers of its
two components (simulations were performed by means of the
split-step algorithm combined with the 2D fast Fourier transform).
The dislocation leads to oscillatory dynamics, and GSs were
classified as stable ones if they would oscillate near the initial
shape. A caveat is that such a definition of the stability does
not make a clear distinction between stable stationary solitons
and stable breathers with a small amplitude of internal
vibrations; however, the objects of both types are, as \ a matter
of fact, experimentally equivalent localized states in the BEC.

Figure \ref{graph_3d_ep10} shows that the stability pattern is more complex
for a stronger OL, with $\varepsilon =10$. In particular, the inter-gap
soliton, which is possible in this case, may be stable when the chemical
potential $\mu _{1}$ of the component with a smaller norm, which appertains
to the first (lower) bandgap, is sufficiently close to the lower edge of the
gap. It is noteworthy that the stability region of the inter-gap solitons is
found at values of the total norm $N$ at which intra-gap solitons, with both
components sitting in the first bandgap, do not exist, i.e., there is no
overlap between these two types of the stable GSs .

\begin{figure}[p]
\includegraphics{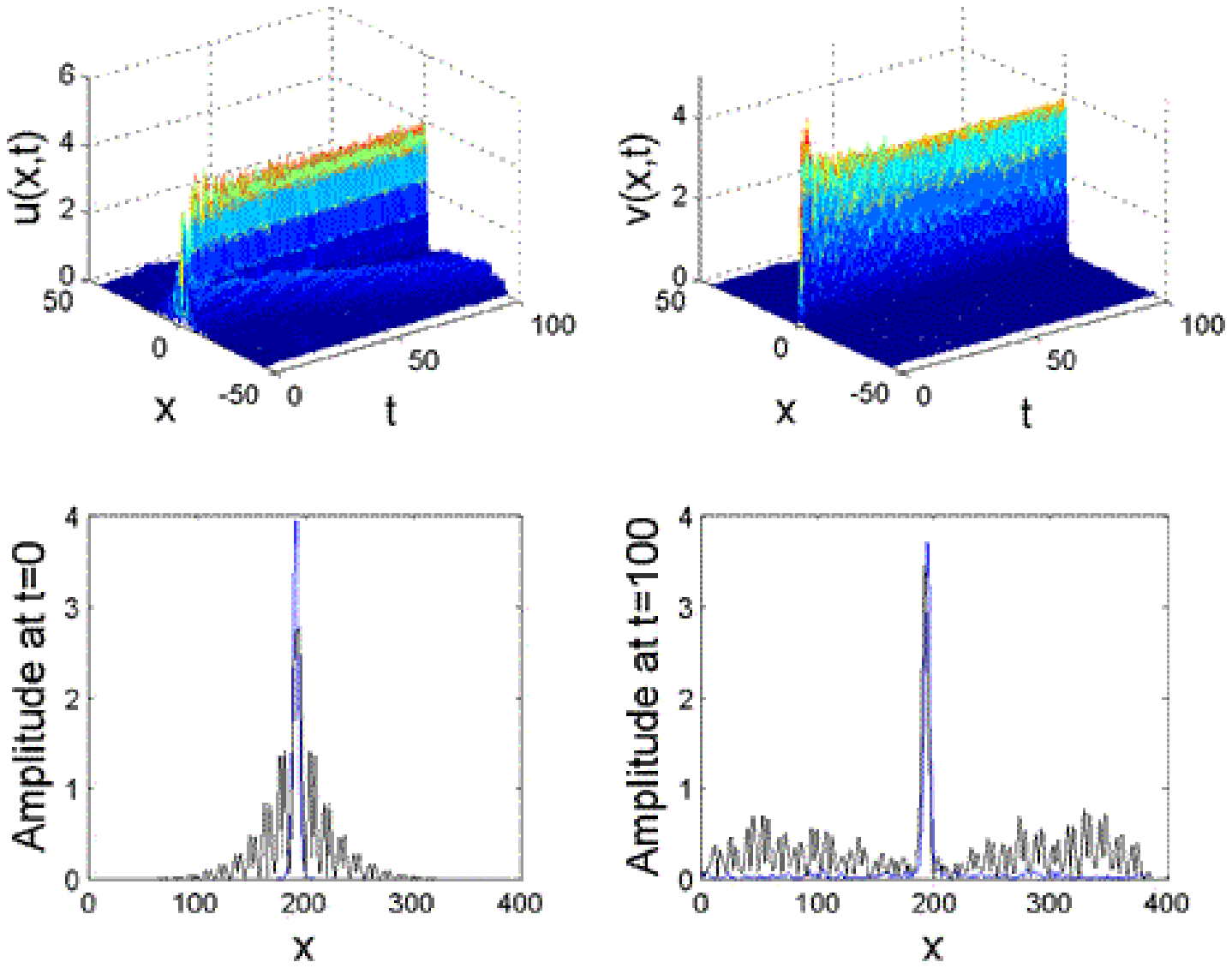}
\caption{(Color online) Evolution of an unstable two-dimensional inter-gap
soliton with $\protect\mu _{1}=-3.7$, $\protect\mu _{2}=-5.6$ is shown in
the cross-section along the $y=0$ axis for $\protect\varepsilon =10$. The
loosely-bound component reduces its norm through emission of radiation.
Eventually, the solution evolves into a symmetric intra-gap soliton of the
tightly-bound type.}
\label{evolve2sym}
\end{figure}

Further simulations demonstrate that unstable inter-gap solitons are either
completely destroyed or evolve into symmetric solitons with both components
belonging to the first gap. An example of such evolution is displayed in
Fig. \ref{evolve2sym}. In this case, the LB component of the TB-LB pair
emits radiation until only the central lobe remains in it, and the whole
structure turns into a symmetric intra-gap soliton of the TB type.

\begin{figure}[p]
\subfigure{\includegraphics{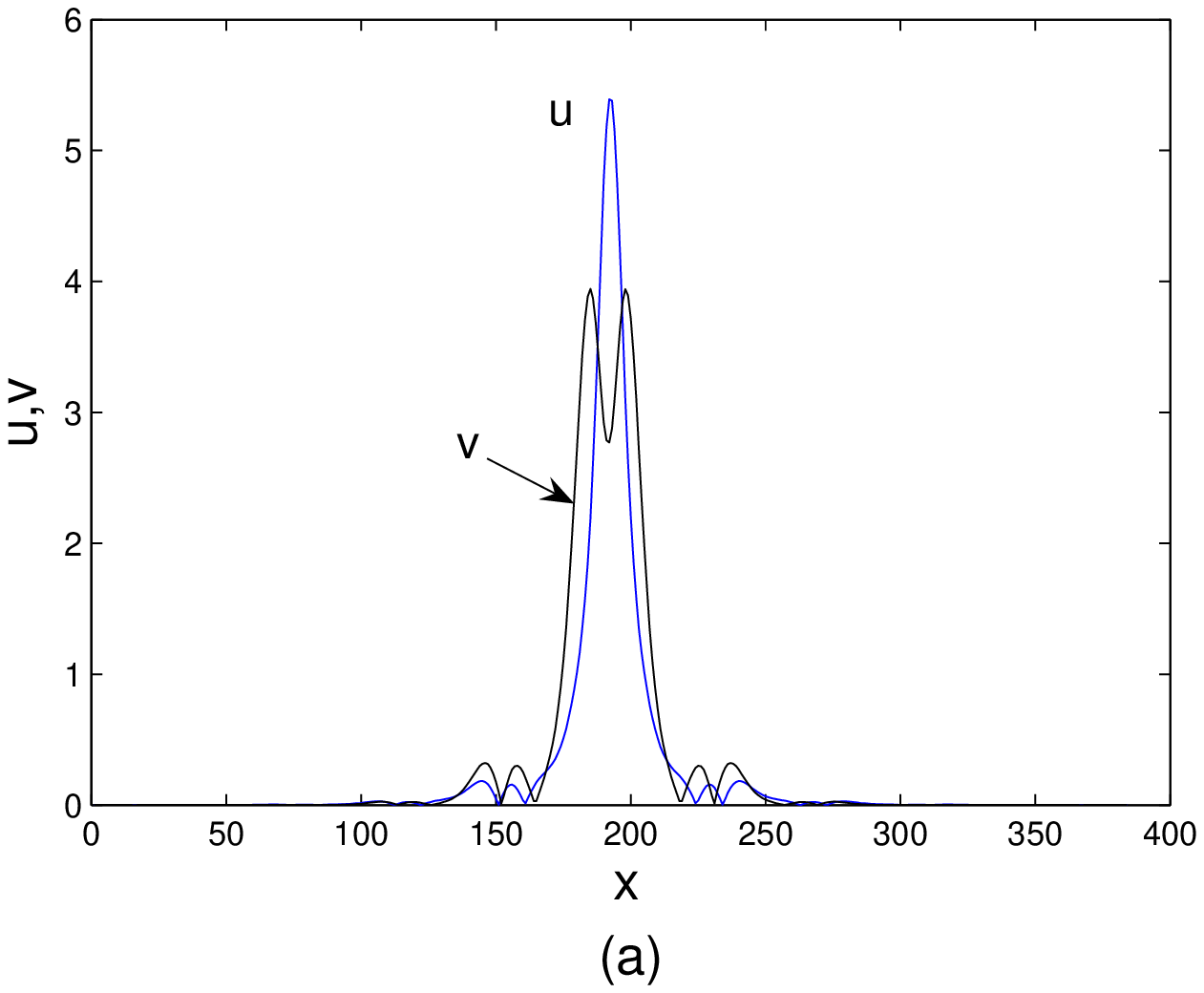}}
\newline \subfigure{\includegraphics{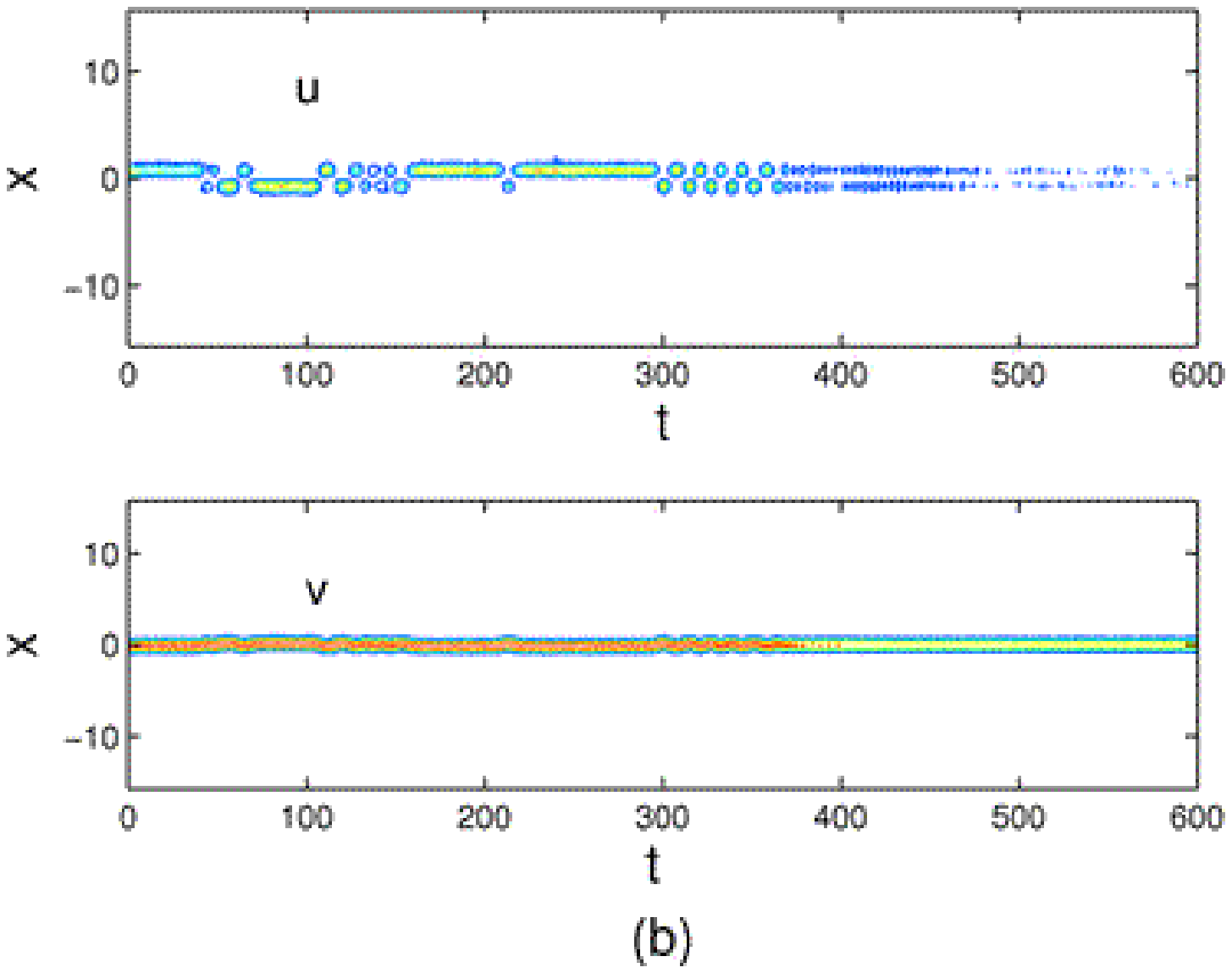}}
\caption{(Color online) The onset of the ``jumping" instability of
the two-dimensional intra-gap soliton with $\protect\mu _{1}=0.5$
and $\protect\mu _{2}=-2.5$, which belongs to the second bandgap,
for $\protect\varepsilon =10$. The upper panel shows the
unperturbed shape of the soliton in the $y=0$ cross section, and
the lower panels display, by means of contour plots, the evolution
of both components in the same cross section.} \label{jumper}
\end{figure}

A specific instability mode was observed in unstable inter-gap solitons, as
well as in intra-gap ones with both components belonging to the second gap,
in the case when the unperturbed soliton features a single peak in one
component and a double (split-tip) peak in the other. In this configuration,
the instability triggers oscillatory dynamics with the single peak jumping
irregularly between positions close to the two side peaks of the mate
component. An example of this instability is displayed in Fig. \ref{jumper}.
Further evolution leads to complete destruction of both components in this
case.

Finally, simulations of the full model (\ref{model}) with $\rho
\neq 0$, which includes the self-repulsion in each component,
reveal a clear trend to stabilization of the two-component GSs as
$\rho $ increases (of course, the model with $\rho >0$ gives rise
to ordinary single-component GSs too). For example, the unstable
symmetric soliton with $\mu _{1}=\mu _{2}=1.2$, $\varepsilon =2$,
and $\rho =0$ (recall the system has a single finite bandgap in
this case) becomes stable as the self-repulsion coefficient
increases to $\rho _{\min }\approx 2/3$. This stabilization
mechanism will be considered in more detail below in the 1D
version of the model.

\section{One-dimensional solitons}

The 1D variant of the model corresponds to a ``cigar-shaped"
binary condensate, tightly confined in the transverse plane, with
a one-dimensional OL created in the longitudinal direction, $x$.
The accordingly modified 1D version of the normalized equations
(\ref{model}) is\begin{eqnarray} i\psi _{t}+\psi _{xx}+\varepsilon
\cos (2x)\psi -\left( \rho |\psi
|^{2}+|\varphi |^{2}\right) \psi  &=&0,  \notag \\
&&  \label{model1D} \\
i\varphi _{t}+\varphi _{xx}+\varepsilon \cos (2x)\varphi -\left( \rho
|\varphi |^{2}+|\psi |^{2}\right) \varphi  &=&0.  \notag
\end{eqnarray}The conserved norms of the scaled wave functions are
\begin{equation}
N_{1,2}\equiv \int_{-\infty }^{+\infty }\left\{ \left\vert \psi
(x)\right\vert ^{2},\left\vert \varphi (x)\right\vert ^{2}\right\} dx,
\label{N12-1D}
\end{equation}which are related to the numbers of atoms as follows:\begin{equation}
\left( N_{\mathrm{phys}}\right) _{1,2}=\frac{k}{4a}\frac{\left(
\int_{0}^{\infty }\chi _{0}^{2}(R)RdR\right) ^{2}}{\int_{0}^{\infty }\chi
_{0}^{4}(R)RdR}N_{1,2},  \label{numbers1D}
\end{equation}cf. Eq. (\ref{numbers}). Here $\pi /k$ is, as above, the OL period,
and $\chi _{0}(R)$ is the ground-state wave function of the tight confining
potential, $R$ being the radial coordinate in the transverse plane.

A principal difference of the 1D model from its 2D counterpart is that, at
any finite $\varepsilon $, the 1D version of the operator (\ref{L}), which
is the same as in the Mathieu equation, gives rise to an infinite system of
finite bandgaps (recall that the 2D operator generates a single finite gap
for small $\varepsilon $, two gaps for larger $\varepsilon $, etc.).

Stationary soliton solutions to Eqs. (\ref{model1D}) are again looked for in
the form of Eqs. (\ref{stationary}) (with the coordinate $y$ dropped). In
contrast to the above analysis of the 2D model, in the present case we
determine the stability of solitons in a rigorous way, from linearized
equations for small perturbations about the stationary soliton (however, in
all the cases when GSs were predicted to be stable in this sense, their
stability was also verified in direct simulations). The application of the
rigorous approach to the 2D model is to be presented elsewhere, as it is a
technically involved problem.

The perturbed solutions are taken as
\begin{eqnarray}
\psi  &=&\left[ u(x)+\psi _{1}(x)e^{-i\lambda t}\right] e^{-i\mu _{1}t},
\notag \\
&&  \label{pert} \\
\varphi  &=&\left[ v(x)+\varphi _{1}(x)e^{-i\lambda t}\right] e^{-i\mu
_{2}t},  \notag
\end{eqnarray}where $\psi _{1}$ and $\varphi _{1}$ are eigenmodes of infinitesimal
perturbations and $\lambda $ is the respective eigenfrequency, the
instability corresponding to having $\mathrm{Im~}\lambda >0$. The
substitution of this in Eqs. (\ref{model1D}) and linearization lead to the
equations
\begin{eqnarray}
&&\left( \mu _{2}+\frac{d^{2}}{dx^{2}}\right) \psi _{1}+\varepsilon \cos
(2x)\psi _{1}-\left[ v^{2}(x)\psi _{1}+u(x)v(x)\left( \phi _{1}+\phi
_{1}^{\ast }\right) \right]   \notag \\
-\rho u^{2}(x)\left( 2\psi _{1}+\psi _{1}^{\ast }\right)  &=&\lambda \psi
_{1},  \notag \\
&&\left( \mu _{1}+\frac{d^{2}}{dx^{2}}\right) \phi _{1}+\varepsilon \cos
(2x)\phi _{1}-\left[ u^{2}(x)\phi _{1}+u(x)v(x)\left( \psi _{1}+\psi
_{1}^{\ast }\right) \right]   \notag \\
-\rho v^{2}(x)\left( 2\phi _{1}+\phi _{1}^{\ast }\right)  &=&\lambda \phi
_{1},  \notag
\end{eqnarray}which can be solved by means of known numerical methods, to yield a full
spectrum of the eigenfrequencies $\lambda $ (we used the Matlab
eigenvalue-finding routine; it is based on approximating the ordinary
differential equations by a system of linear homogeneous algebraic
equations, computing the determinant of the corresponding matrix, and
equating it to zero, which eventually leads to an equation for $\lambda $).

\begin{figure}[p]
\subfigure{\includegraphics[width=3in]{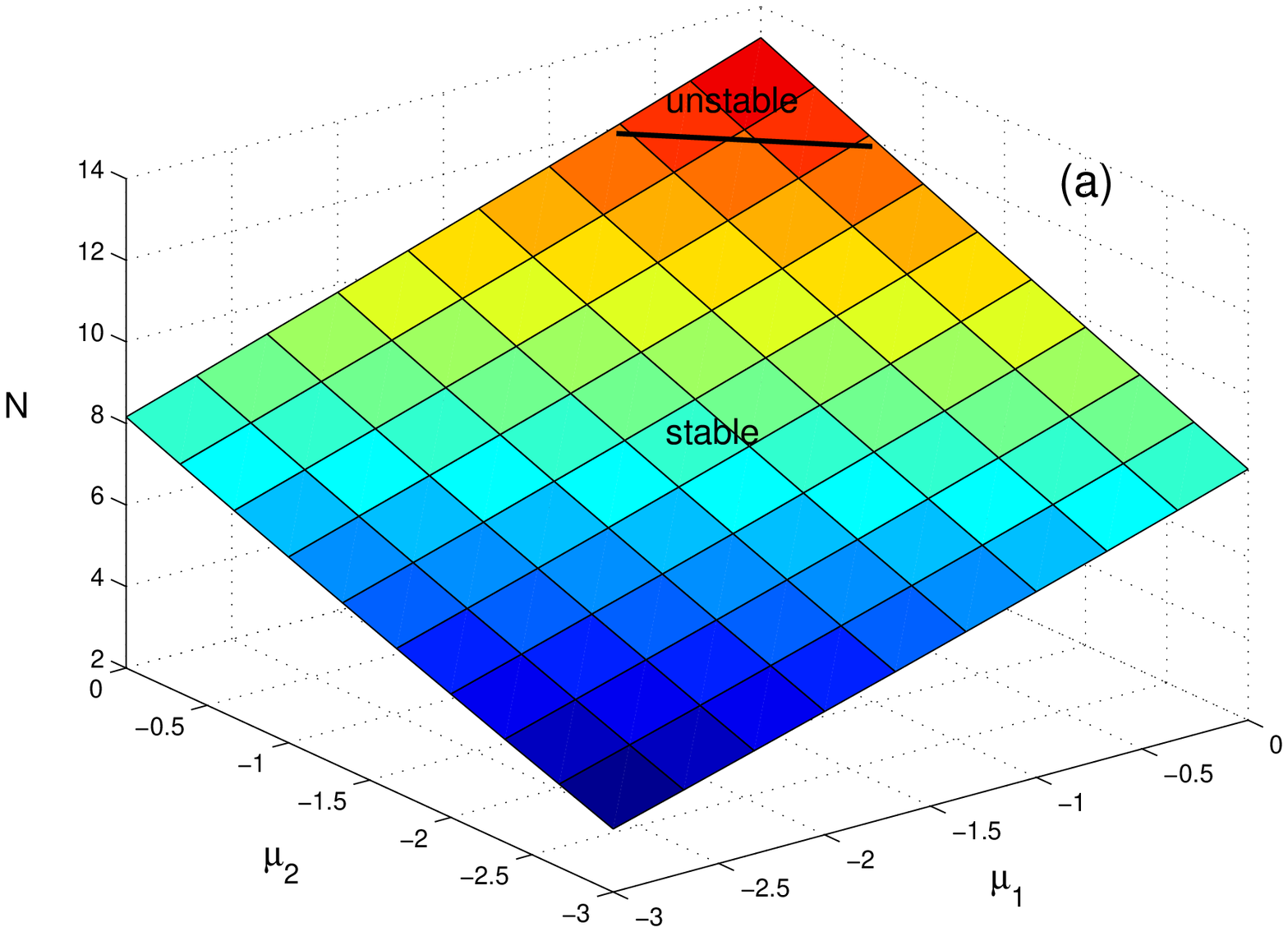}}
\subfigure{\includegraphics[width=3in]{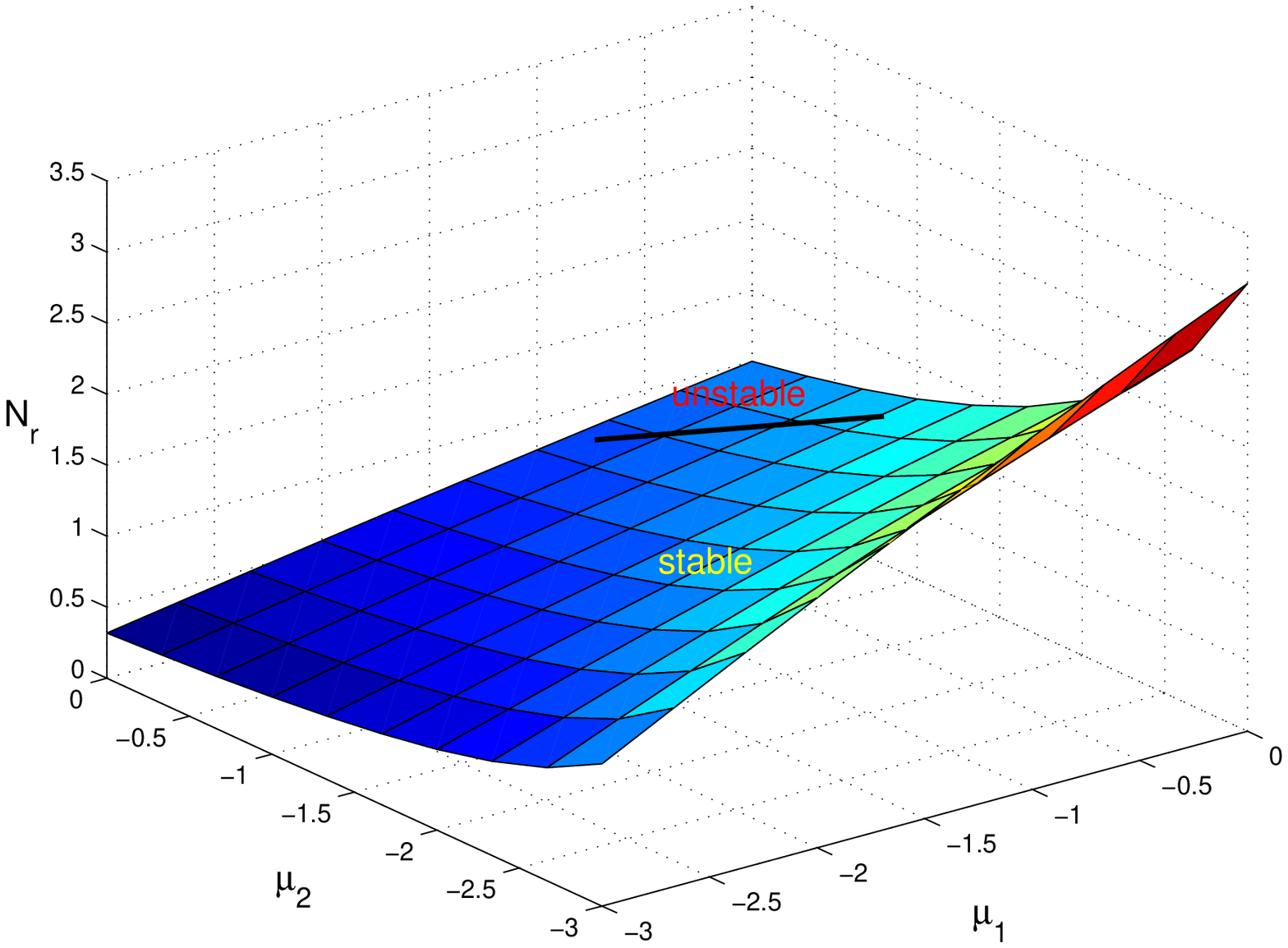}}
\newline
\subfigure{\includegraphics[width=3in]{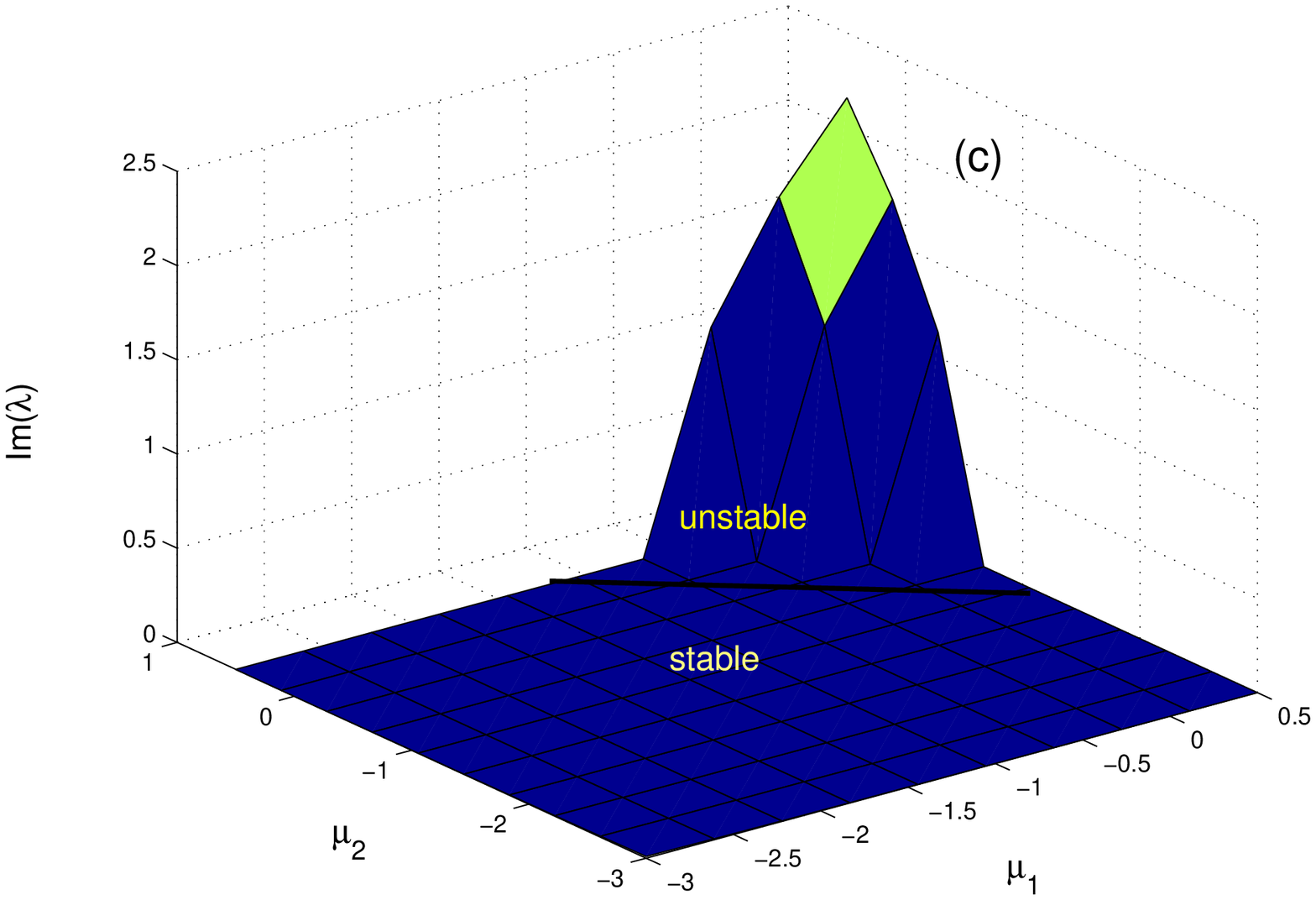}}
\subfigure{\includegraphics[width=3in]{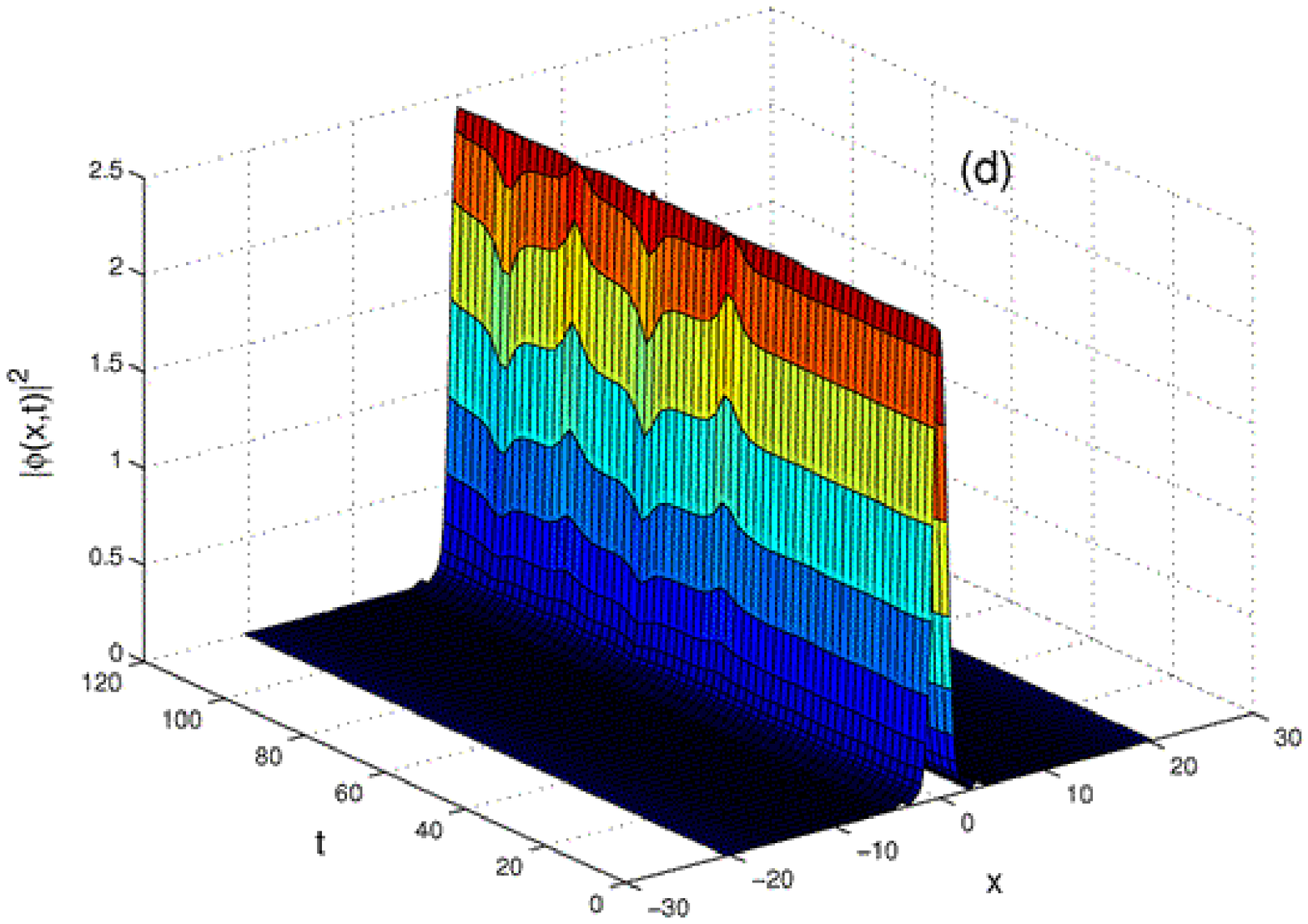}}
\caption{(Color online) A typical example of a family of two-component intra-gap
solitons in the one-dimensional model, with both components
belonging to the first finite bandgap. In this case,
$\protect\varepsilon =8$ (and $\protect\rho =0$). Panels (a), (b),
and (c) show, respectively, the total norm $N=N_{1}+N_{2}$,
relative norm $N_{r}=N_{1}/N_{2}$, and instability growth rate,
$\mathrm{Im~}\protect\lambda $. Panel (d) displays an example of
the transformation of an unstable symmetric intra-gap soliton
(with $\protect\mu _{1}=\protect\mu _{2}=0$) into a stable
breather, due to the oscillatory instability.} \label{Fig0Ilya}
\end{figure}

First of all, we present results for the most fundamental case of
$\rho =0$, when only two-component GSs are possible, as well as in
the 2D model considered above. Fixing the OL strength $\varepsilon
$, in Figs. \ref{Fig0Ilya}(a) and (b) we display the total and
relative norms, $N$ and $N_{r} $ for the family of 1D
two-component GSs, as functions of the two chemical potentials,
$\mu _{1}$ and $\mu _{2}$, for a case when both belong to the
first finite bandgap (i.e., for the family of intra-gap solitons)
[the norms $N$ and $N_{r}$ are defined as in Eqs. (\ref{NN}), with
$N_{1}$ and $N_{2}$ taken as per Eq. (\ref{N12-1D})]. The
stability of the same GS family is presented in panel (c) of Fig.
\ref{Fig0Ilya}, which shows the largest value of
$\mathrm{Im~}\lambda $, i.e., the instability growth rate, as a
function of $\mu _{1}$ and $\mu _{2}$ [the border between stable
and unstable solitons is indicated in Figs. \ref{Fig0Ilya}(a) and
(b) too]. If the GS of this type is unstable, its instability is
oscillatory (i.e., $\mathrm{Im~}\lambda >0$ comes along with
$\mathrm{Re~}\lambda \neq 0$). Typically, the instability does not
destroy the soliton, but rather transforms it into quite a stable
breather, see an example in Fig. \ref{Fig0Ilya}(d).

\begin{figure}[p]
\subfigure{\includegraphics[width=3in]{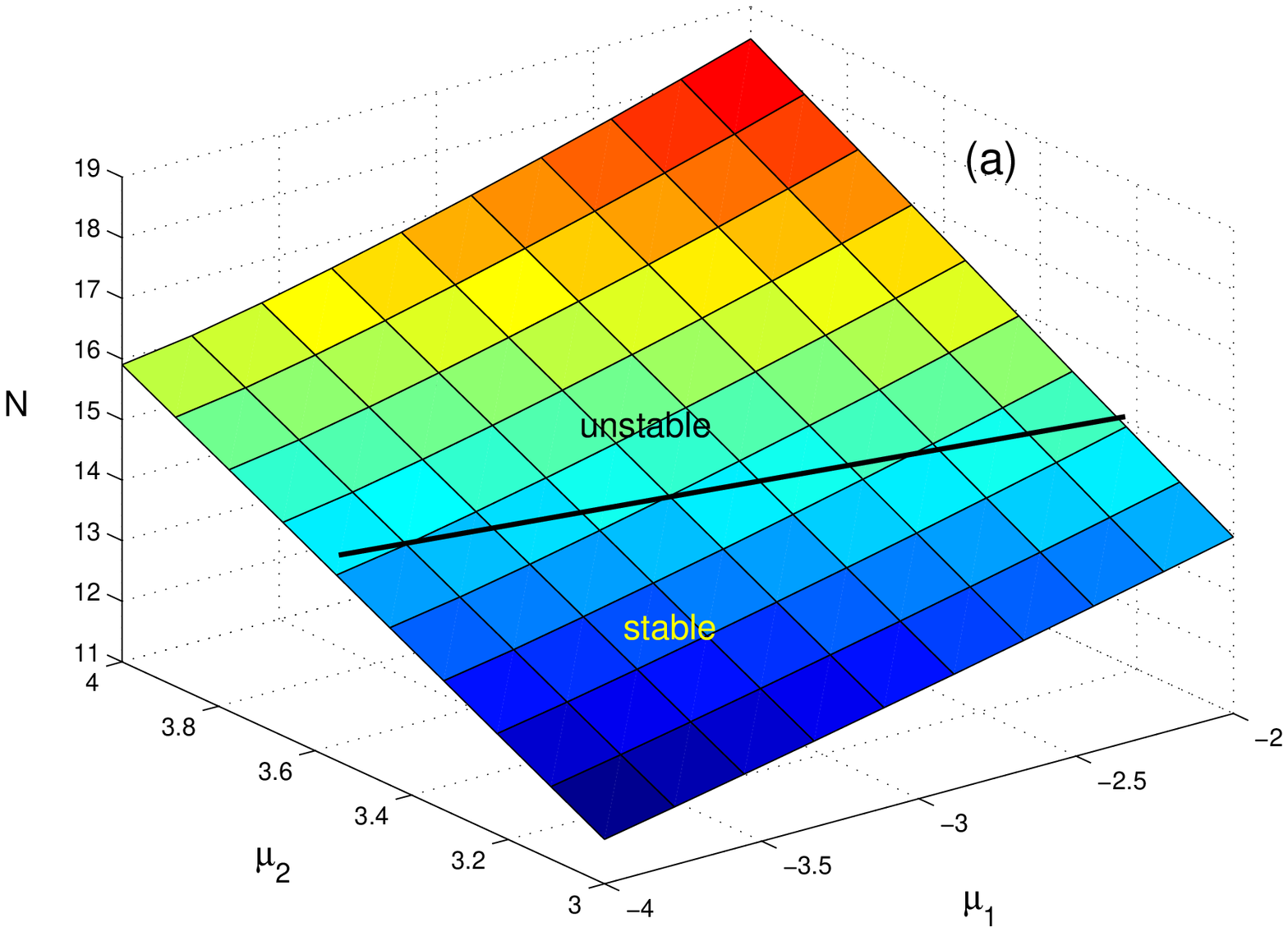}}
\subfigure{\includegraphics[width=3in]{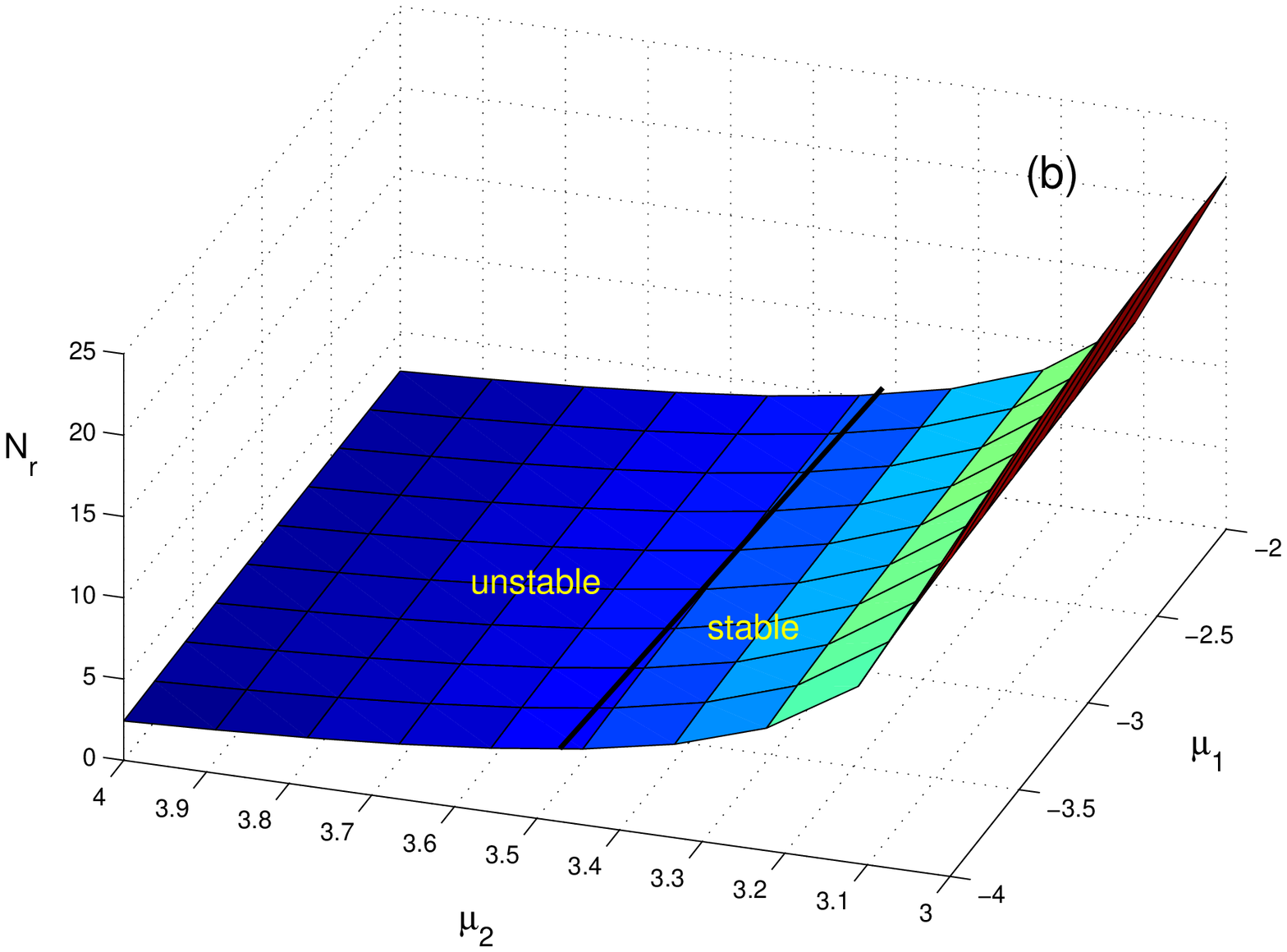}}
\newline
\subfigure{\includegraphics[width=3in]{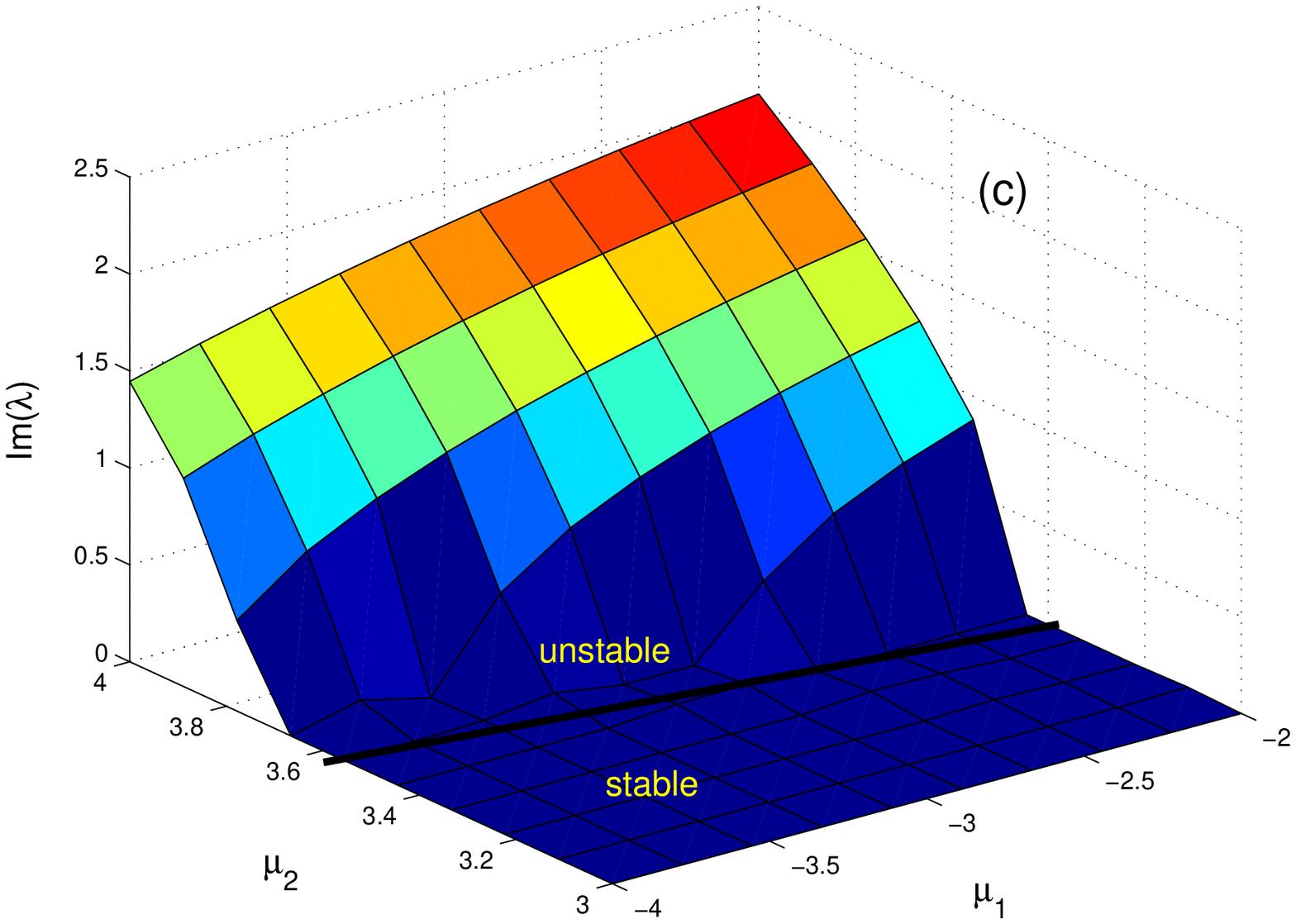}}
\subfigure{\includegraphics[width=3in]{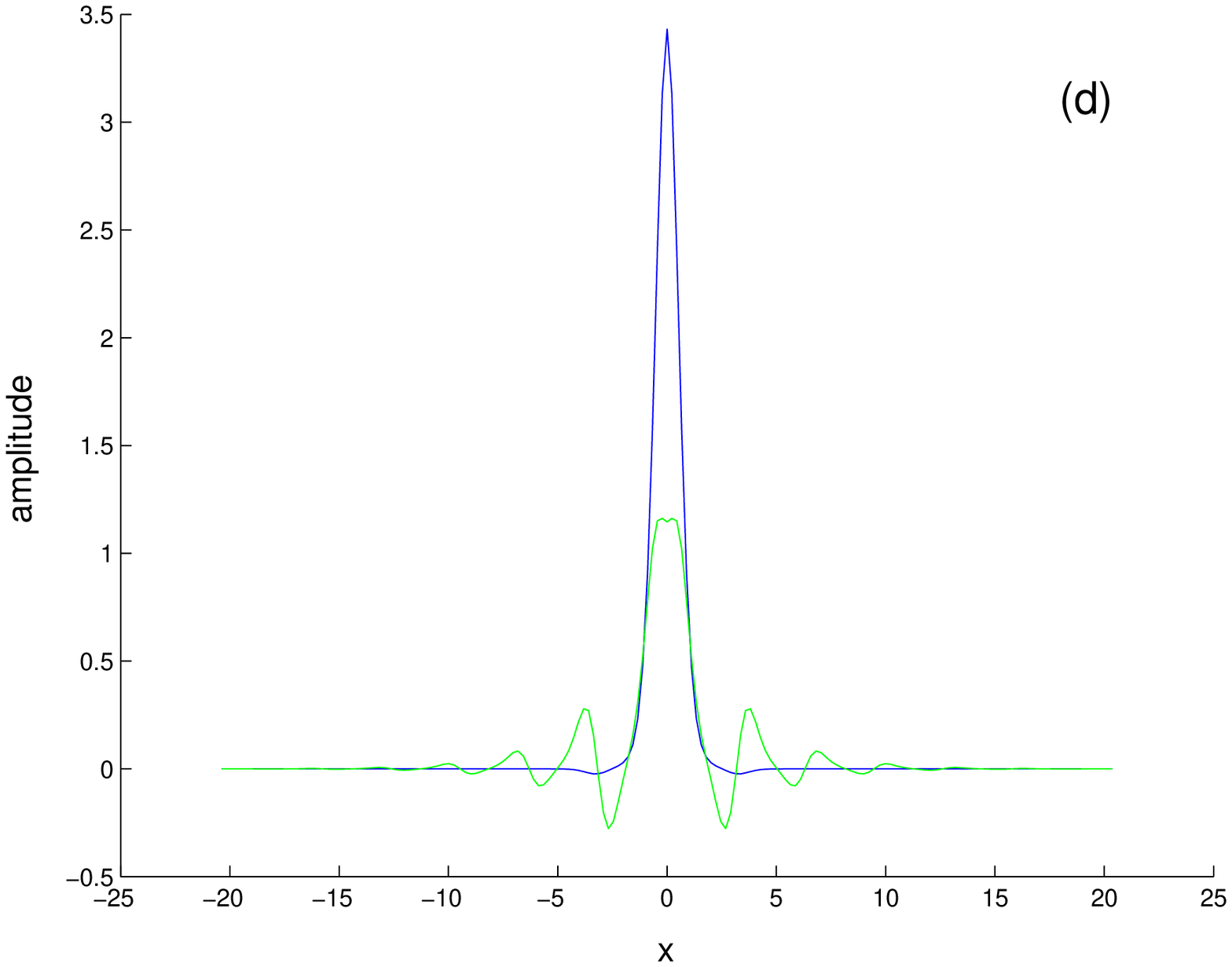}}
\caption{(Color online) Panels (a), (b), and (c) show the same
characteristics as in Fig. \protect\ref{Fig0Ilya} for a typical
family of two-component inter-gap solitons in the one-dimensional
model, with one component belonging to the first finite bandgap,
and the other -- to the second. In this case, $\protect\varepsilon
=8$ and $\protect\rho =0$. Panel (d) displays a generic example of
a stable inter-gap soliton, with $\protect\mu _{1}=-3$,
$N_{1}=11.7$, and $\protect\mu _{2}=3$, $N_{2}=0.75$.}
\label{Fig1Ilya}
\end{figure}

Inter-gap solitons, built of components belonging to the fist and
second finite bandgaps, have also been investigated in detail in
the 1D model. Characteristics of this solution family are
presented in Fig. \ref{Fig1Ilya}. As seen from panel (c) of the
figure, the stability area of the inter-gap solitons is
essentially smaller than in the case of their intra-gap
counterparts, cf. Fig. \ref{Fig0Ilya}(c). Note that panels (a),
(b) and (c) in Fig. \ref{Fig1Ilya} do not show a Bloch band that
separates the two gaps (unlike Fig. \ref{graph_3d_ep10} in the 2D
model), as the ranges of values on the axes of $\mu _{1}$ and $\mu
_{2}$ in these panels cover, respectively, only the first and
second gaps.

It is noteworthy that, as well as in the 2D model, the inter-gap
solitons are bound states of TB and LB components belonging to the
first and second gaps, respectively. An example of a stable
inter-gap soliton that clearly demonstrates this structure is
shown in Fig. \ref{Fig1Ilya}(d) [cf. Figs.
\ref{va_vs_num_n50_ep10}(d) and (e) and Fig. \ref{num_ep4}, which
display cross sections of intra-gap solitons in the 2D model].

\begin{figure}[tbp]
\centering\subfigure{\includegraphics[width=3in]{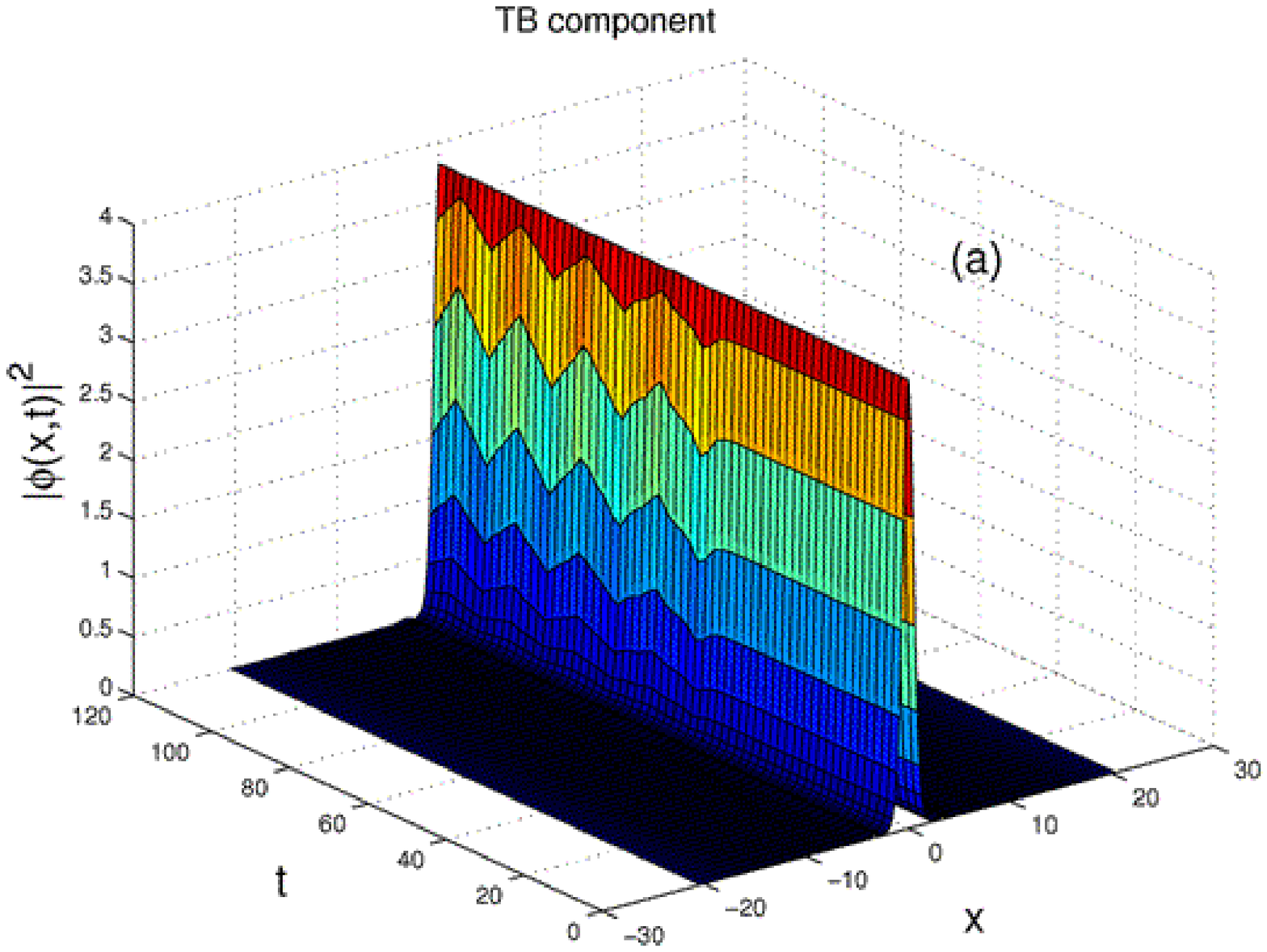}}
\centering\subfigure{\includegraphics[width=3in]{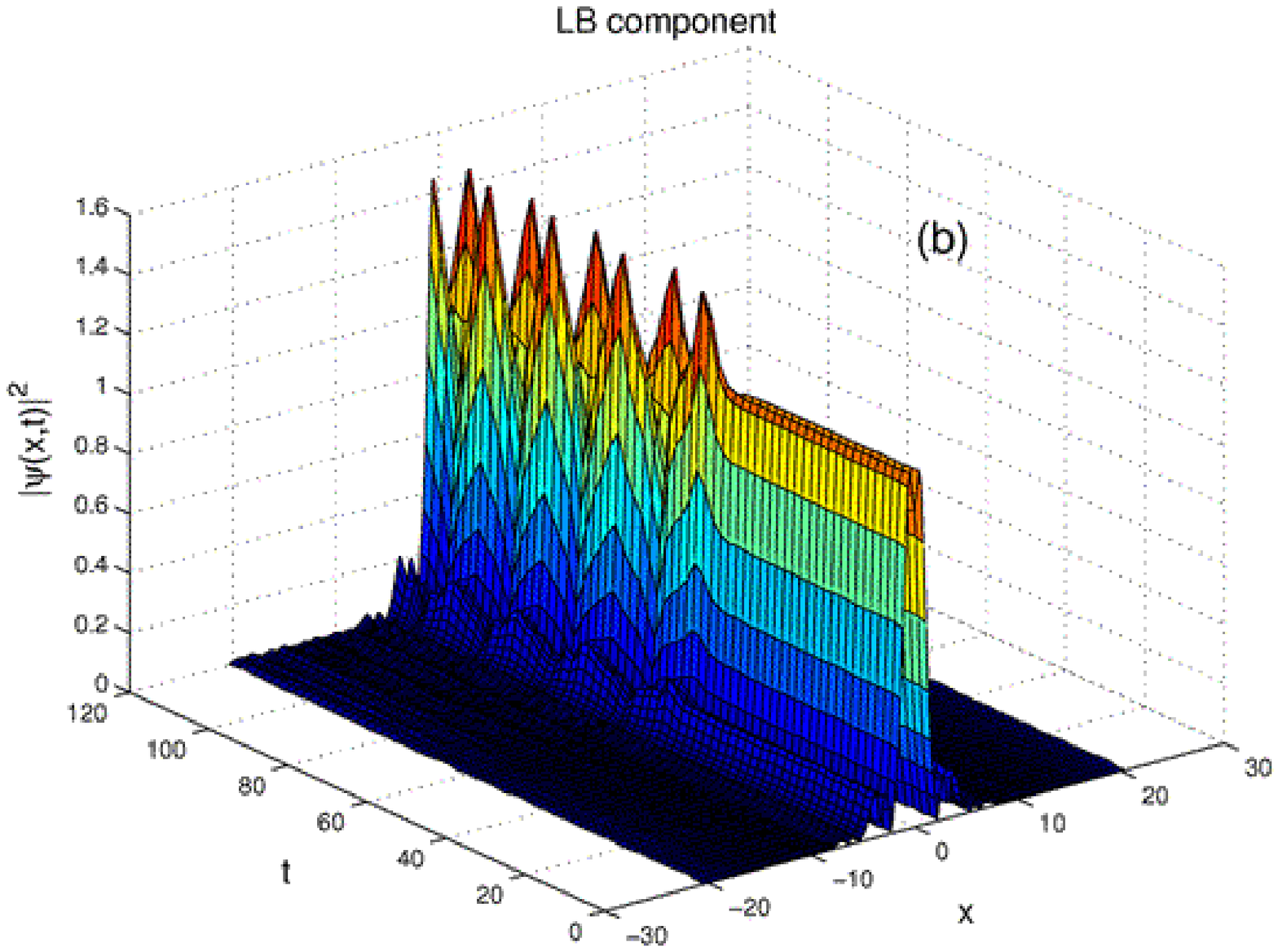}}
\caption{(Color online) An example of the evolution of an unstable
inter-gap soliton in the one-dimensional model with
$\protect\varepsilon =8$, $\protect\rho =0$ and $\protect\mu
_{1}=-3$, $\protect\mu _{2}=4$. Panels (a) and (b) show the
development of the instability in the components that originally
belong to the first and second finite gaps, respectively.}
\label{Fig2Ilya}
\end{figure}

If an inter-gap soliton is unstable, its instability again has the
oscillatory character. However, the action of the instability on the
inter-gap soliton is more destructive than it was in the case of its
intra-gap counterpart: instead of transforming the stationary soliton into a
well-localized breather [see Fig. \ref{Fig0Ilya}(d)], the instability
triggers much more violent evolution, as illustrated by a typical example in
Fig. \ref{Fig2Ilya}.

\begin{figure}[tbp]
\centering{\includegraphics[width=3in]{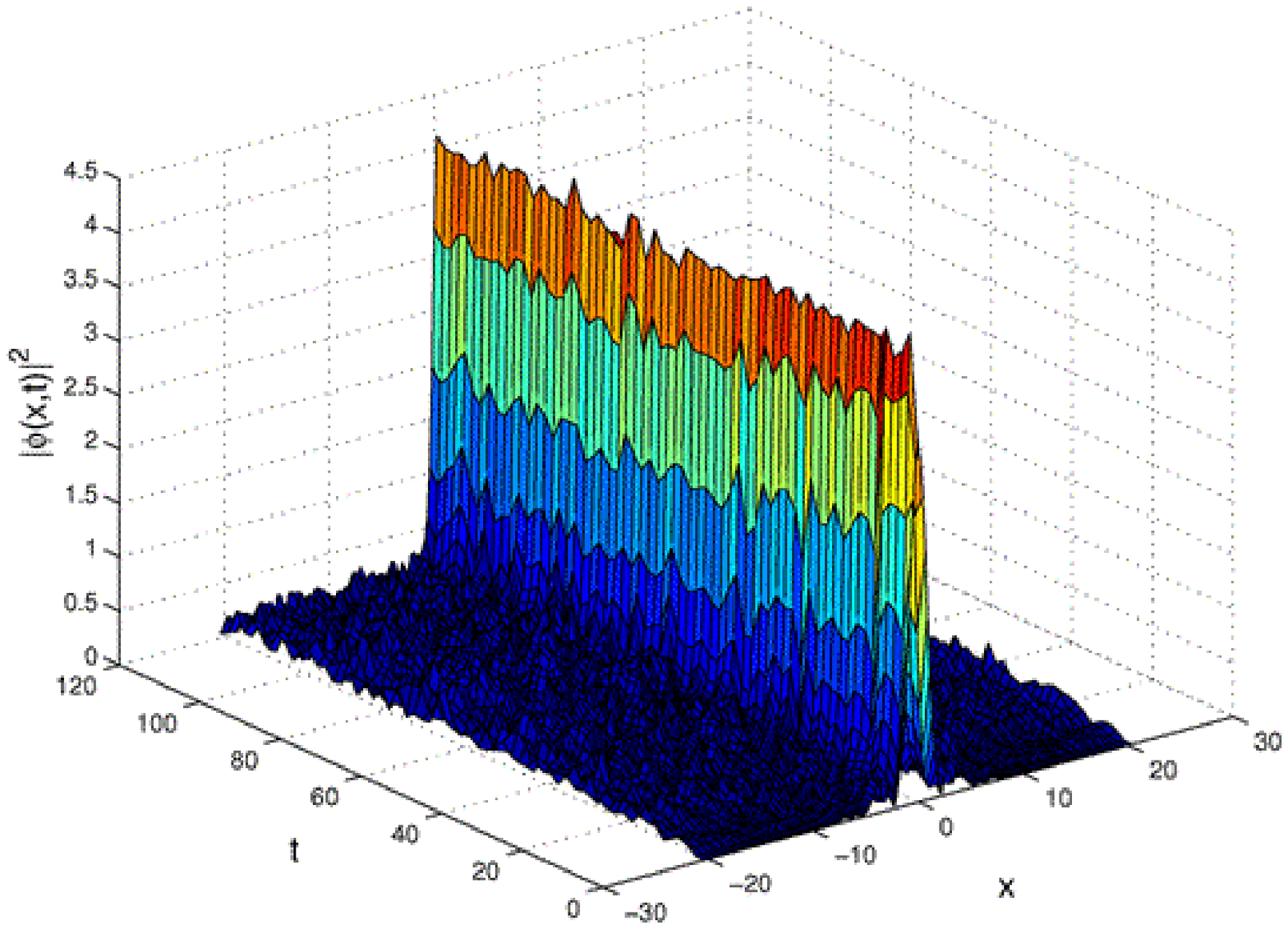}}
\caption{(Color online) Instability of a symmetric intra-gap soliton
appertaining to the second finite bandgap in the one-dimensional model. In
this case, $\protect\varepsilon =8$, $\protect\rho =0$ and $\protect\mu =4$.
}
\label{Fig3Ilya}
\end{figure}

One-dimensional intra-gap solitons belonging to the second gap
were constructed and investigated too, with a conclusion that they
all are \emph{unstable}, although the instability growth rate may
sometimes be very small. An example illustrating the instability
of this species of the GS is displayed in Fig. \ref{Fig3Ilya}.
This conclusion seems to be in contrast with results reported
above for the 2D model, where a small stability area for the
intra-gap solitons appertaining to the second finite bandgap was
found, see Fig. \ref{graph_3d_ep10}. However, the stability of the
2D solitons was identified not via eigenfrequencies of small
perturbations, but rather by means of direct simulations, and the
numerical stability test always turned the stationary soliton into
a weakly excited state (a breather with a small amplitude of
intrinsic vibrations). Therefore (as it was said above), in the 2D
case we, strictly speaking, cannot tell a difference between a
weakly unstable stationary soliton and a weakly excited stable
breather into which the dynamical evolution may transform the
unstable soliton. Thus, it may happen that what was identified as
stable 2D intra-gap solitons sitting in the second finite bandgap
are, actually, stable breathers with a small vibration amplitude,
similar to what is observed in Fig. \ref{Fig0Ilya}(d).

\begin{figure}[tbp]
\centering\includegraphics[width=4in]{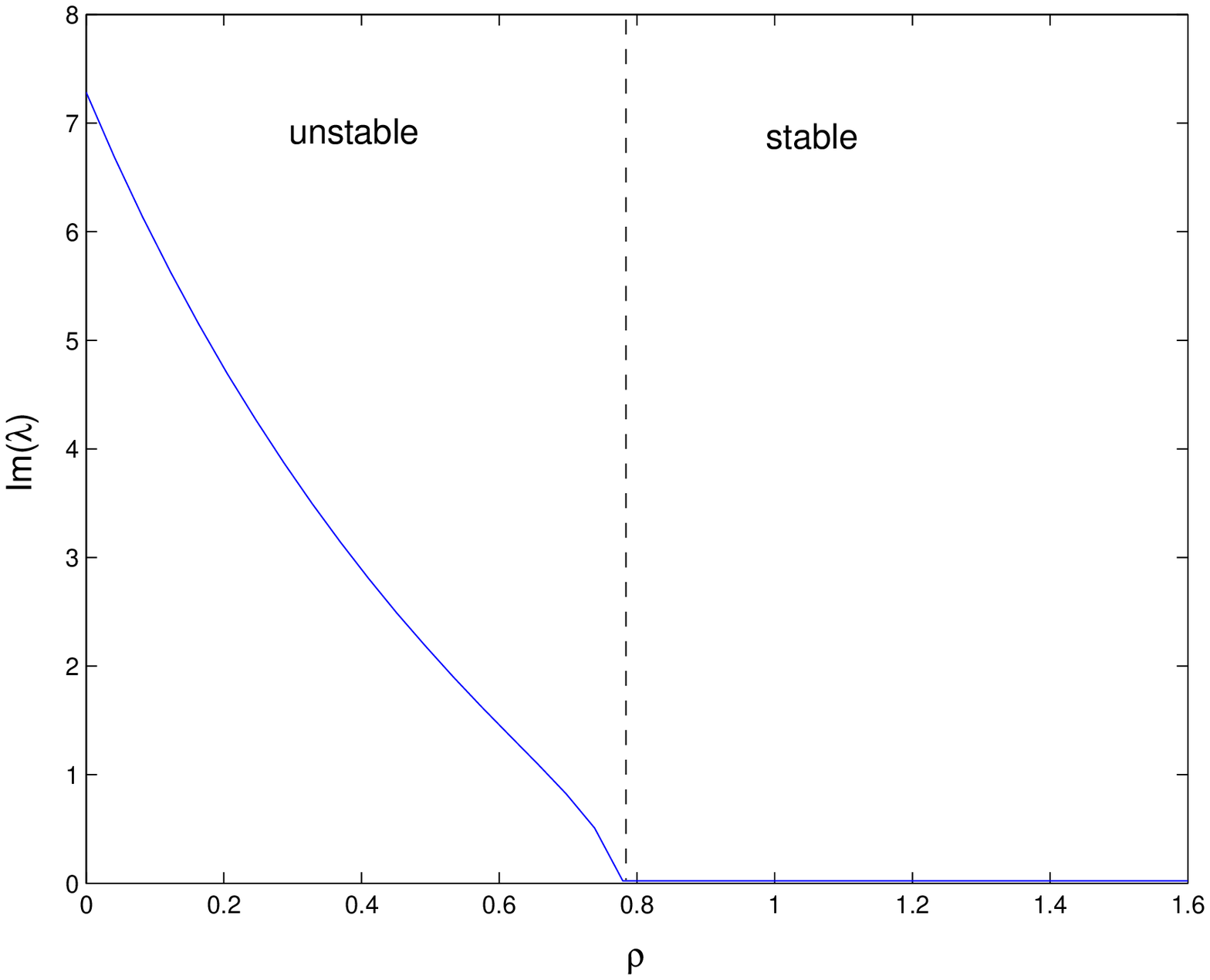}
\caption{(Color online) The instability growth rate of a symmetric
soliton in the second finite bandgap vs. the self-repulsion
coefficient $\protect\rho $ [see Eqs. (\protect\ref{model1D})].
Except for $\protect\rho $, parameters are the same as in the
previous figure. The soliton is stable in the region of
$\protect\rho >\protect\rho _{\min }\approx 0.8$.}
\label{Fig4Ilya}
\end{figure}

We have also investigated in some detail the effect of the self-repulsion
term in the full 1D system (\ref{model1D}). As was already mentioned above
in connection to the 2D model, the increase of the self-repulsion
coefficient $\rho $ leads to stabilization of the GSs. A similar effect in
the 1D setting is clearly seen in Fig. \ref{Fig4Ilya}. It demonstrates that
a stable symmetric intra-gap soliton belonging to the second finite bandgap,
which, as said above, is always unstable in the 1D model with $\rho =0$,
becomes stable if $\rho $ exceeds a minimum (threshold) value, which, in
this case, is $\rho _{\min }\approx 0.9$.

The above stability analysis was performed within the framework of
the GPE, i.e., in the mean-field approximation at zero
temperature. A physically important issue is stability of the same
solitons against quantum and thermal fluctuations beyond the
framework of the mean-field theory. The interest to this issue is
enhanced by experimental observation of quite strong effective
re-thermalization the ultra-cold condensate under the action of
the OL potential \cite{thermo}. To extend the analysis in this
direction, one may use a system of self-consistent time-dependent
Hartree-Fock-Bogoliubov (TDHFB)\ equations, built around the
corresponding Gross-Pitaevskii equation(s) \cite{TDHFB}. This
approach was recently used to demonstrate possible instability of
an ordinary 1D soliton (not of the gap type) in the BEC with
attractive interactions, which is completely stable in the
framework of the GPE proper \cite{MotiAvi}. In that case, the
instability splits the soliton into two segments.

A properly modified system of the TDHFB equations (including the OL
potential and repulsive inter-species interaction) can also be used for the
investigation of the extended stability of the two-component GSs studied in
this work. We have performed a preliminary analysis along these lines, and
concluded that those 1D solitons which are stable within the framework of
Eqs. (\ref{model1D}) are also stable (at least, in most cases) against
fluctuations governed by the TDHFB equations. A systematic consideration of
this issue, especially for the 2D model, requires a considerable amount of
additional work and will be reported elsewhere.

\section{Conclusion}

In this work, we have introduced a model of a binary BEC with intrinsic
inter- and intra-species repulsion (positive scattering lengths), which is
loaded into the periodic optical-lattice potential. Both two- and
one-dimensional versions of the model were considered. We focused on the
most fundamental case (different from previously studied models), with
repulsion between the two species and zero intra-species interaction, which
can be achieved by means of the Feshbach-resonance (FR)\ technique, or in a
spinor condensate. The same system may also model a mixture of two mutually
repulsive fermionic species.

The main problem was the existence and stability of gap solitons (GSs)
supported by the interplay of the inter-species repulsion and periodic
potential. Two-component GSs were looked for by means of the variational
approximation (VA) and in the numerical form. It was found that the VA
provides for good accuracy in the case when the 2D soliton is a bound state
of two tightly-bound components, each being essentially confined around one
cell of the periodic potential. Such a GS structure dominates in the case
when both components belong to the first finite bandgap of the system's
spectrum (the intra-gap soliton). In fact, only this type of the GS is
possible in a weak 2D lattice potential. On the contrary, inter-gap
solitons, and intra-gap ones residing in the second finite bandgap (both
types are possible in a stronger lattice potential, with the barrier height
essentially larger than the recoil energy) are, typically, bound states of
tightly- and loosely-bound components. For such structures, the VA is
irrelevant, but general results were obtained in the numerical form, which
made it possible to identify the existence and stability regions for the
inter- and intra-gap solitons in both the 2D and 1D models. In the 2D case,
the stability was tested in direct simulations, while in the 1D model the
stability was identified in a rigorous way, through the computation of
eigenfrequencies of small perturbations (the results were also verified by
direct simulations). In the case when the 1D intra-gap soliton belonging to
the first gap is weakly unstable, it evolves into a stable breather with a
small amplitude of intrinsic vibrations. In contrast to this, if the 2D
solitons in the first gap are unstable, they are completely destroyed by the
instability. The same pertains to unstable 1D and 2D solitons of other
types, such as inter-gap solitons, and intra-gap ones belonging to the
second finite bandgap. It was also shown that introduction of the
intra-species repulsion, in addition to the repulsion between the
components, leads to further stabilization of solitons. In particular, some
originally unstable types (such as 1D intra-gap solitons in the second
bandgap) may be made stable, provided that the self-repulsion coefficient
exceeds a certain minimum value. Preliminary analysis shows that the
two-component GSs introduced in this work are stable too against
fluctuations obeying the Hartree-Fock-Bogoliubov equations.

The actual number of atoms in the 2D solitons considered above, which is
their most important physical characteristic, can be easily estimated.
Undoing renormalizations which led to the 2D Gross-Pitaevskii equations in
the form of Eqs. (\ref{model}) and making use of Eq. (\ref{numbers}), one
can derive the following relations between the density of atoms in physical
units, $n_{\mathrm{phys}}$, and the rescaled one $|\psi |^{2}$, and between
the number of atoms and the soliton's norm:
\begin{equation}
n_{\mathrm{phys}}=\frac{\pi }{2a_{s}\Lambda ^{2}}|\psi
|^{2},~N_{\mathrm{phys}}=\frac{l_{z}}{8\pi a_{s}}N,
\label{physics}
\end{equation}where $\Lambda $ is the period of the optical lattice, $a_{s}$ the
scattering length of the inter-species collisions, and $l_{z}$ the
size of the condensate in the transverse ($z$) direction (note
that the relation between $N_{\mathrm{phys}}$ and $N$ does not
contain $\Lambda $). Assuming experimentally reasonable values,
$a_{s}\sim 1$ nm and $l_{z}\simeq 2$ $\mu $m (the latter pertains
to a ``pancake-shaped" condensate trapped between two
diffraction-limited blue-detuned repelling laser sheets
\cite{pancake}), we conclude that typical values, $N\simeq 20$ and
$N\simeq 100$, for the stable intra-gap and inter-gap solitons,
respectively [see Figs. \ref{graph_3d_ep2}(a) and
\ref{graph_3d_ep10}(a)], lead to the following estimates for the
respective numbers of atoms: $N_{\mathrm{intra}}\sim 5,000$,
$N_{\mathrm{inter}}\sim 25,000$. If necessary, this number may be
made at least an order of magnitude larger, reducing $a_{s}$ by
means of the FR technique \cite{inter-Feshbach}. It is relevant to
mention that, in the first observation of a GS in BEC
\cite{Oberthaler}, the number of atoms $\simeq 1,000$ in the
soliton's central lobe was quite sufficient for the experiment.

For the effectively 1D condensate in a cigar-shaped trap
\cite{Pethick}, Eq. (\ref{numbers1D}) leads to the following
equation replacing the second relation in Eq.
(\ref{physics}),\begin{equation}
N_{\mathrm{phys}}=\frac{A}{4a_{s}\Lambda }N,  \label{1D}
\end{equation}where $A$ is an effective transverse area of the trap [note that this
relation, unlike its 2D counterpart in Eqs. (\ref{physics}), explicitly
contains the optical-lattice period $\Lambda $]. Taking the same $a_{s}$ as
above, $\sim 1$ nm, a typical $\Lambda \sim 1$ $\mu $m, and a reasonable
value of $A\simeq 30$ $\mu $m$^{2}$ (it corresponds to the effective trap's
radius $\simeq 3$ $\mu $m), we conclude that, with the characteristic value
of the norm for the stable intra- and inter-gap 1D solitons, $N\simeq 10$ in
the normalized units [see Figs. \ref{Fig0Ilya}(a) and \ref{Fig1Ilya}(a)],
Eq. (\ref{1D}) yields an estimate $\sim 10^{5}$ for the actual number of
atoms in the effectively 1D soliton.

\section*{Acknowledgements}

We appreciate valuable discussions with V. P\'{e}rez-Garc\'{\i}a, M.
Salerno, and A. Vardi. This work was supported, in a part, by the Israel
Science Foundation through the Excellence-Center grant No. 8006/03.

\end{document}